\documentclass{article}
\usepackage{graphicx}
\usepackage{hyperref}
\usepackage{amsmath}

\title{Analyzing gender inequality through large-scale Facebook advertising data}

\author
{David Garcia,$^{a,b}$ Yonas Mitike Kassa,$^{c,d}$\\Angel Cuevas,$^{d}$ Manuel Cebrian,$^{e,f}$\\Esteban Moro,$^{g}$ Iyad Rahwan,$^{f,h}$ Ruben Cuevas$^{d}$
\vspace{0.25cm}
\\
\normalsize{$^{a}$ Complexity Science Hub Vienna, Vienna, Austria}\\
\normalsize{$^{b}$ Medical University of Vienna, Vienna, Austria}\\
\normalsize{$^{c}$ IMDEA Networks Institute, Madrid, Spain}\\
\normalsize{$^{d}$ Universidad Carlos III de Madrid, 28911 Leganes, Spain}\\
\normalsize{$^{e}$ Data61, CSIRO, Melbourne, Australia}  \\
\normalsize{$^{f}$ The Media Lab, Massachusetts Institute of Technology, Cambridge MA, USA}\\
\normalsize{$^{g}$ Department of Mathematics \& GISC,}\\\normalsize{Universidad Carlos III de Madrid, 28911 Leganes, Spain}\\
\normalsize{$^{h}$ Institute for Data, Systems \& Society,} \\\normalsize{Massachusetts Institute of Technology, Cambridge MA, USA}
}

\begin{document}

\pagestyle{headings}
\markright{Please cite PNAS version:  \url{https://doi.org/10.1073/pnas.1717781115}}

\maketitle

\begin{abstract}

Online social media are information resources that can have a transformative power in society.
While the Web was envisioned as an equalizing force that allows everyone to access information, the digital divide prevents large amounts of people from being present online.
Online social media in particular are prone to gender inequality, an important issue given the link between social media use and employment.
Understanding gender inequality in social media is a challenging task due to the necessity of data sources that can provide large-scale measurements across multiple countries.
Here we show how the Facebook Gender Divide (FGD), a metric based on aggregated statistics of  more than 1.4 Billion users in 217 countries, explains various aspects of worldwide gender inequality.
Our analysis shows that the FGD encodes gender equality indices in education, health, and economic opportunity.
We find gender differences in network externalities that suggest that using social media has an added value for women.
Furthermore, we find that low values of the FGD are associated with increases in economic gender equality.
Our results suggest that online social networks, while suffering evident gender imbalance, may lower the barriers that women have to access informational resources and help to narrow the economic gender gap.
\end{abstract}

The Web was designed to be universally accessible and open, carrying the promise of equal opportunity in the access to online information and services \cite{Berners2010} as \emph{the great potential equalizer} \cite{Hargittai2013}.
However, despite the widespread adoption of the Web and other Information Communication Technologies (ICT), online access is heterogeneously distributed across demographic factors, such as income and gender --a phenomenon called the \emph{digital divide} \cite{Brown1995,Compaine2001,Norris2001}.


Governments and global organizations express their concern about the digital divide, aiming to connect the 4 Billion people that remain offline \cite{UN2006,Deichmann2016}.
However, the effects of increasing Internet penetration in development are rarely backed up against empirical data \cite{Friederici2017}, and the latest report by the World Bank suggests that unequally distributed growth in Internet penetration might exacerbate socio-economic inequalities \cite{Deichmann2016}.
Beyond the divide in the access to Internet, there are further challenges with respect to digital inequality: the heterogeneity of online activity and engagement across demographic groups \cite{Hargittai2013}.

Among online resources, social media play a key role in economic development, for example by providing information that facilitates finding employment \cite{Burke2013,Gee2017}.
An important open question is whether equality in the access to social media can work as a digital provide \cite{Jensen2007}, bringing equality in other social, political, and economic aspects of society.
The World Wide Web Foundation reports that one of the key elements in the digital divide is gender inequality \cite{www2015}.
Social media data show the traces of gender inequalities, from content biases and activity on Wikipedia \cite{Wagner2016,Rizoiu2016} to visibility and interaction disparities on Twitter \cite{Garcia2014,Nilizadeh2016,Magno2014} and 
professional gender gaps in LinkedIn \cite{Haranko2018}.
Empirical analyses of digital traces has the potential to track more general demographic patterns \cite{Billari2017,Fatehkia2018}, such as fertility rates \cite{Ojala2017}.




To properly understand the digital divide, a pervasive problem in cross-country comparisons is the limited size of country samples and the challenges to generate unbiased survey data \cite{Deichmann2016}.
To overcome this issue, we deployed a system to collect large-scale data from the Facebook online social network through its marketing API, as explained more in detail in the Materials and Methods section.
The Facebook marketing API has been useful in previous research to estimate the value of user data \cite{Gonzalez2017}, to approximate the size and integration of migrant populations \cite{Potzschke2017,Zagheni2017,Dubois2018}, and to generate estimations of Internet and mobile phone gender gaps that explain 69\% of the variance of ITU measurements \cite{Fatehkia2018}.
For our study of the relationship between social media gender divides and other economic, education, heath, and political gender inequalities, we generated an anonymous dataset with statistics about the total number of registered users and daily active users of each gender in each country. While our dataset does not contain personal information on any individual user, our study covers a total of 217 countries and more than 1.4 Billion users.


Our dataset allows the quantification of Facebook activity ratios of each gender in each country.
From them, we calculate the \emph{Facebook Gender Divide} (FGD) as the logarithm of the ratio between the activity ratios for men and for women (see Methods for more details). The FGD has a value below zero when women tend to be more active on Facebook that men, a value close to zero for equal activity tendencies, and a positive value when men are more active on Facebook than women in a country. Our computation of the FGD is consistent with similar measurements constructed from limited survey samples from the Pew Research Center and the Global Web Index, as we comment in the Methods section and show in Supplementary Text 1.

Furthemore, the Facebook marketing API allows us to make precise estimates of the \emph{Facebook penetration} in a country, calculated as the total number of user accounts (independent of gender and activity) over the total population of the country. We combine these new measurements with standard socio-economic indices, including Gross Domestic Product (GDP), Internet penetration, and economic inequality, as well as indices from the World Economic Forum Gender Gap Report that measure gender equality in terms of education, health, political participation, and economic opportunities \cite{WEF}.

\begin{figure}
\centering
\includegraphics[width=\linewidth]{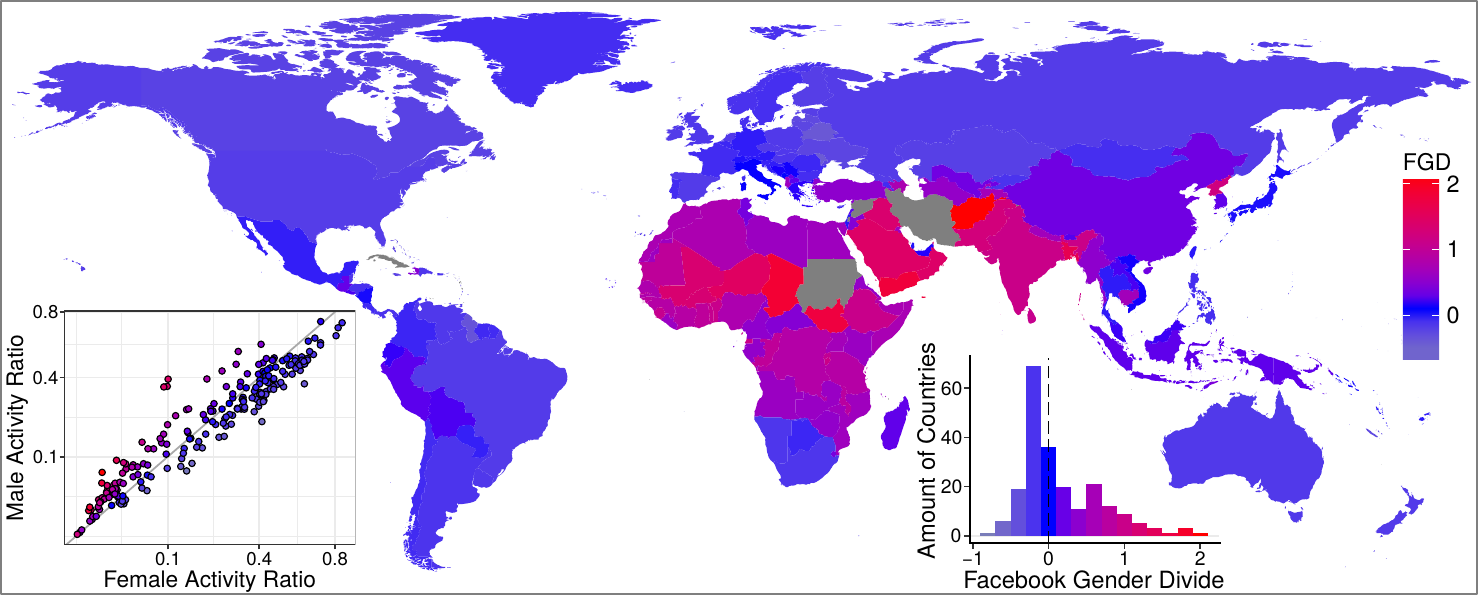}
\caption{\textbf{The Facebook Gender Divide across 217 countries.} Countries are colored according their Facebook Gender Divide (FGD), from highly skewed towards males (red), balanced (blue), and towards females (green, not visible). The left inset shows the scatter plot of male and female activity ratios across all countries, revealing a spread along the diagonal. The right inset shows the histogram of FGD values in bins of width 0.2. While the mode of countries is slightly below zero, there is significant skewness towards high FGD values. An online interactive version of this figure can be found in \url{https://dgarcia-eu.github.io/FacebookGenderDivide/Visualization.html}.}
\label{fig:frog}
\end{figure}

\section*{Results}



Fig. 1 shows a world map with countries colored according to their FGD, revealing that many countries are very close to gender equality in Facebook (blue color).
The red scale shows countries with positive FGD--that is, higher proportion of males on Facebook. The range of values towards FGD below zero (more tendency for women to be on Facebook) is much narrower than above zero, as can be seen in the scatter plot with the activity ratios of each gender (Fig. 1, left inset), and in the skewness of the distribution of FGD across countries (Fig. 1, right inset).


Countries with high FGD are located around Africa and South West Asia, as shown on Fig. 1.
This suggests that variations in socio-economic factors of gender inequality across regions could be explanatory of the FGD.
We test this observation using a linear regression model of the FGD as a function of the four indices of gender equality measured by the World Economic Forum (economic opportunity, education, health, and political participation), plus five non-gender-based controls of Internet penetration, population size, economic inequality, Facebook penetration, and mean Facebook active user age  (see Methods). The left panel of Fig. 2 shows the quality of the model fit, comparing empirical values of FGD rank versus model predictions. Remarkably, the model can explain well the ranking of FGD ($R^2 = 0.74$), with very few points far from the diagonal. While this result might be partially explained by Facebook using vital statistics in their calculations, it is nevertheless consistent with replications of the model using limited survey samples from the Pew Research center and the Global Web Index (see Supplementary Text 2). This indicates that the performance of the model is not an artifact of the Facebook marketing API.

\begin{figure}
\centerline{\includegraphics[width=\textwidth]{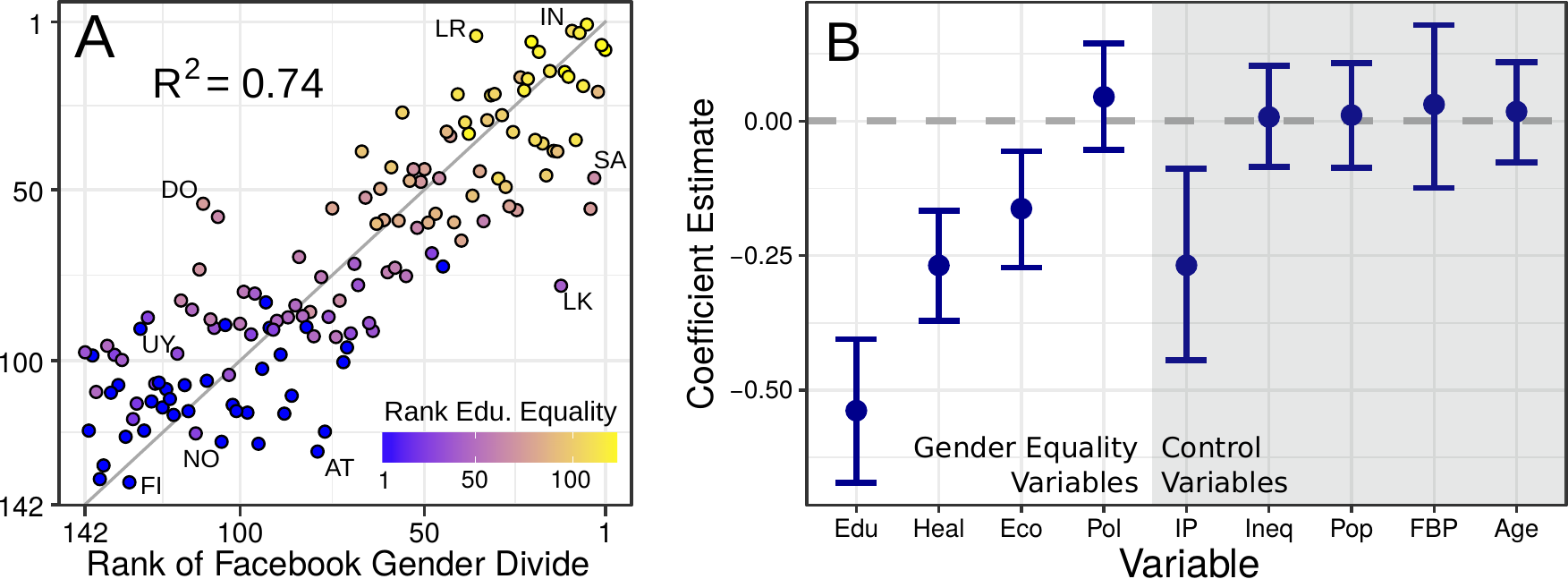}}
\caption{\textbf{Regression results of FGD as a function of gender equality.} A) Model predictions versus rank of FGD, where rank 1 is the country with the highest FGD. The model achieves a high $R^2$ above $0.74$, explaining the majority of the variance of the FGD ranking.
Some countries are labeled, from high FGD (Liberia, India, and Saudi Arabia) to low FGD (Finland, Norway, and Uruguay), as well as some outliers (Dominican Republic, Austria, and Sri Lanka). B) Coefficient estimates and 95\% CI of the terms of the regression fit (excluding intercept). Education (Edu), health (Heal), and economic gender equality (Eco) are significantly and negatively associated with the FGD, but political gender equality (Pol) is not. From the control variables, Internet penetration (IP) is negatively associated with FGD, but the rest are not. The main role of education equality in FGD can be observed on panel A, where dots are colored according to the rank of education gender equality, showing that countries with low FGD are ranked high on education gender equality. An online interactive version of this figure can be found in \url{https://dgarcia-eu.github.io/FacebookGenderDivide/Visualization.html}}
\end{figure}

The right panel of Fig. 2 shows the estimate of the coefficients of our model of FGD. 
The strongest coefficient is that of education gender equality, which can also be observed on the colors of the left panel of Fig. 2. Specifically, countries with high rank in this index have, on average, lower FGD. Health and economic gender equality also have significant negative coefficient estimates, showing that the FGD captures more than one type of inequality.
Note that the index for political gender equality does not have a significant relationship with FGD when the other indices are considered in the model.

Among gender-independent controls, only Internet penetration is negatively associated with FGD.
Nevertheless, the FGD is also correlated with GDP per capita (Spearman correlation $-0.57$ , $p < 10^{-6}$).
For that reason, we repeated the model using GDP as a control variable, finding similar results.
These results evidence that the relationship between gender equality indices and FGD is observable when development metrics are considered.
We present these additional controls, regression diagnostics, and robustness tests in Supplementary Text 2, concluding that the negative relationships between FGD and gender equality indices are robust.


The value of being active in social media might vary across genders, which we address in a wide country comparison.
The general penetration of a communication channel can increase the value that individuals get for using it, which is an example of a  feedback mechanism driven by (positive) \emph{network externalities} \cite{Katz1985}, also known as  Metcalfe's law \cite{Hendler2008}.
If there are network externalities on Facebook, the activity ratio of countries should scale superlinearly with the Facebook penetration in each country.
This scaling relationship with Facebook penetration might vary for the activity ratios of different genders, which would signal an additional marginal benefit of using Facebook for one gender.


Fig. 3 shows the scaling relationship per gender between the activity ratio and the total Facebook penetration in each country. Lines show the result of a power-law fit between both variables with an intercept and an interaction term for gender. The estimate of the scaling exponent for each gender is clearly above one for both genders, revealing a superlinear trend consistent with network externalities in Facebook. This exponent is significantly stronger for female users ($\alpha_F=1.45$ CI=$[1.41,1.49]$) than for male users ($\alpha_M=1.20$ CI=$[1.16,1.24]$, see Supplementary Text 3 for details), suggesting that the network externalities in Facebook are stronger for women than for men.

\begin{figure}
\centerline{\includegraphics[width=\textwidth]{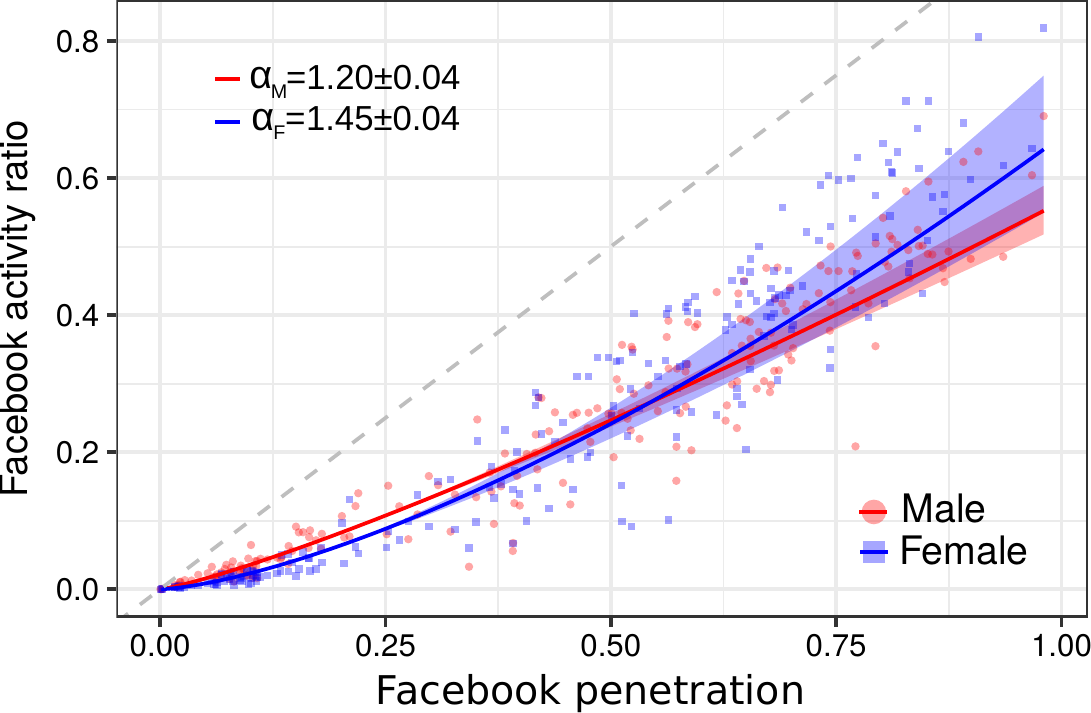}}
\caption{\textbf{Gender differences in network externalities on Facebook.} 
Scaling of Facebook activity ratio per gender versus total Facebook penetration. Solid lines show fit results and shaded areas show their 95\% confidence intervals. Both male and female activity ratios grow superlinearly with Facebook penetration ($\alpha>1$), indicating positive network externalities. These network externalities are stronger for female than for male users ($\alpha_F>\alpha_M$).}
\end{figure}


Given the network externalities shown above, could the FGD be related to changes in economic gender inequality?
We test this possibility by analyzing the change in FGD and Economic gender equality between 2015 and 2016. 
We fitted two regression models, one of changes of Economic gender equality as a function of FGD 
($FGD_{2015} \rightarrow \Delta Eco_{2016}$), and the converse one ($Eco_{2015} \rightarrow \Delta FGD_{2016}$), including controls for autocorrelation and GDP as explained in the Methods section.
The coefficient estimates, shown in Fig. 4, reveal a significant positive relationship between the FGD rank and changes in economic gender inequality, but not vice versa: there is no significant relationship between Economic gender equality and the changes in FGD. 

The partial $R^2$ value of $FGD_{2015}$ in the first model is much higher than the equivalent of $Eco_{2015}$ in the second model (median bootstrap values of $0.027$ and $0.002$ respectively), as shown in the second column of Fig. 4. This suggest the existence of an association between FGD and changes in Economic gender equality such that countries with a low value of FGD (i.e. high rank number) tend more on average to approach economic gender equality.
This observation is consistent across age groups and is robust to the inclusion of further control variables including socio-economic indicators, other gender equality metrics, and Hofstede's culture values \cite{Hofstede2003} (see Supplementary Text 4 for more details).
On the contrary, this association is not observable for education gender inequality, as a model of $\Delta Edu_{2016}$ shows no significant coefficient for $FGD_{2015}$.

\begin{figure}
\centerline{\includegraphics[width=\textwidth]{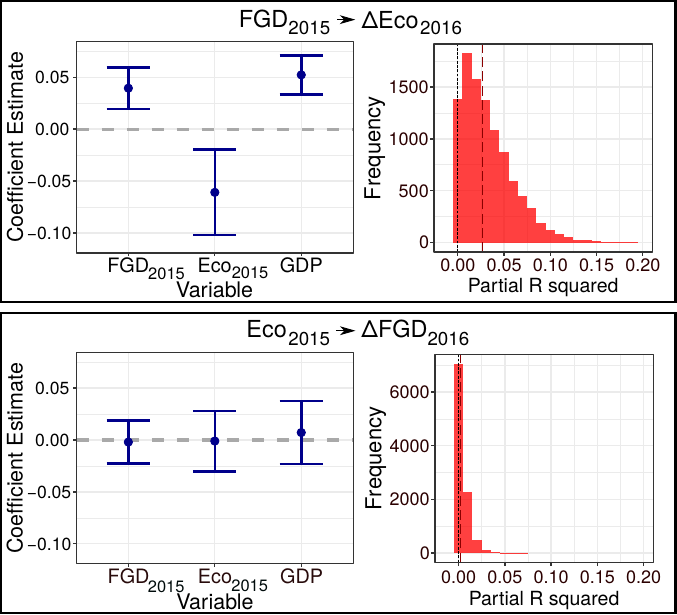}}

\caption{\textbf{Analysis of changes in economic gender equality and FGD.} 
Coefficient estimates of the regression model of changes in economic gender equality as a function of FGD and control terms (excluding intercept, top left), and of the model of changes in FGD as a function of economic gender equality and control terms (excluding intercept, bottom left). Right panels show the bootstrap distributions of partial $R^2$ of $FGD_{2015}$ in the first model and of $Eco_{2015}$ in the second one, with dashed vertical lines showing the median $R^2$ values: $0.027$ in the first model and $0.002$ in the second one. The FGD explains changes in economic gender equality much better than economic gender equality explains changes in the FGD.}

\end{figure}

\section*{Discussion}

By quantifying the Facebook Gender Divide among 1.4 Billion Facebook users, we demonstrate a number of phenomena that deserve further investigation. The FGD is associated with other types of gender inequality, including economic, health, and education inequality.  While the mechanisms behind this connection and its generalizability to other social media remain as open questions, this work is an example of how publicly accessible social media data can be used to understand an important social phenomenon.

Recent reports warn about the possibility that individual Facebook user data was misused by Cambridge Analytica \cite{Ingram2018}, pointing to general concerns about privacy in social media. We share those concerns, in particular with respect to the use of sensitive data in potential conflict with the EU General Data Protection Regulation \cite{Gonzalez2018} and regarding the possible construction of shadow profiles of non-users \cite{Garcia2017,Garcia2018}.
Nevertheless, our results shows that non-personal data, e.g. anonymous and aggregated data produced by billions of Facebook users can be used for social good, in particular to understand the issue of gender inequalities in society at large.

We found evidence of  gender-dependent network externalities---that is, women might receive higher marginal benefit than men from the general adoption of Facebook in a country. While we only observe traces of this phenomenon at an aggregate level, in the differences between activity rates across countries, these results point towards a new research direction: using observational data to understand the value of social networking sites across demographic attributes.

The FGD provides an inexpensive and accessible way to compute gender divides in social media that can be tracked over time and across the vast majority of countries.
This allowed us to identify a relationship between the FGD in 2015 and changes in economic gender inequality in 2016.
This relationship could be produced by three mechanisms: i) a causation path between the FGD and changes in economic gender equality, ii) a more complex causation from economic gender equality on changes in FGD, or iii) by the prevalence of a third factor of cultural gender norms that drive both the FGD and economic gender equality. While we find evidence for the first explanation, we must note that the real interplay between the FGD and economic gender equality is probably a combination of all three mechanisms, and only future research with more detailed data can answer how.

Our results show trends across a wide range of countries, but caution should be taken when extrapolating to the future or when predicating about individual countries. Before doing so, we need longitudinal models of changes in development factors in a wide range of countries, to find the role of the FGD in broader development sequences that include economic development, health, education, and inequality \cite{Spaiser2014} before formulating policy suggestions.
Nevertheless, our results allow us to speculate that social media can be an equalizing force that counteracts other barriers---e.g. those that limit women's mobility \cite{Uteng2012}---by providing access to greater economic opportunities and social capital. In a similar way as mobile phones increased the life quality of fishermen in India \cite{Jensen2007}, social media might work as a \emph{digital provide} that helps disfavored groups despite the still generalized inequalities in access to ICT and in adoption of social media technologies.

\section*{Materials and methods}

\subsection*{The Facebook Global Dataset} 

We collected the number of Facebook users by age and gender in each country using the Facebook marketing API \cite{API}.
Among other services, this API delivers data for its commercial customers to provide targeted advertising.
When supplied with a specific target population, the API returns the total audience size and the price to reach that target audience through Facebook.
We iterated over each combination of age and gender values, retrieving the total number of users and the number of Daily Active Users (DAU) for each segment in each country.
Our dataset contains the number of male and female registered users and DAU for all available countries\footnote{The API does not deliver data for certain countries, e.g. Syria, Iran, and Cuba.}. After removing entries of small countries with missing values or low resolution, our dataset contains the total number of users and DAU segmented by age and gender for 217 countries.
Age data in the API starts at 13, increasing by one year up to a last bin that contains all users aged 65 or older. 
We distribute a dataset to allow the replication and extension of our results through a Github repository\footnote{\url{https://github.com/dgarcia-eu/FacebookGenderDivide}}.

\subsection*{Ethical considerations}

Our analysis of data from the Facebook marketing API only includes aggregated public information.
Even though the sample includes data from underage Facebook users, we had no access to any personal identifiable information of any user and we did not interact or manipulate any research subject. 
The data retrieval was performed as part of the TYPES project funded by the European Comission (GA-653449) and was approved by the Committee of Ethics in Research of the Carlos III University of Madrid (Ethics Report CEI-2015-001).
Our analysis of the data, in line with the growing consensus in ethics \cite{Metcalf2016}, is exempted from ethics review, as agreed by the board of IMDEA Networks and by the executive office of the Complexity Science Hub Vienna.
Nevertheless, following the guidelines of the Association of Internet Researchers \cite{Ess2002}, we consider the possible downstream consequences of our large-scale research.
The resolution of the Facebook marketing API prevents the singling out of individual users, which makes all our codes useless to identify individuals of any minority or threatened group. In addition, there is no way to identify the accounts of users and employ our analysis for any kind of personalization or individual manipulation.
From the onset, our project had the potential to reveal important relationships between social media use and gender inequalities online and offline.
These benefits greatly outweigh the minimum risks of analyzing this kind of aggregated data that is accessible to anyone with an Internet connection.

\subsection*{Validating the FGD} 

Facebook provides the raw data for our study as aggregated values, but as with any new research method, we should not take it at face value without comparing it with more established methods. This is of special importance given the challenges previously found with health-related data from this API \cite{Araujo2017}.

To validate our measurements, we use three reference survey datasets: the Global and Internet \& Technology surveys of the Pew Research Center \footnote{\url{http://www.pewglobal.org/dataset/spring-2016-survey-data/}}\footnote{\url{http://www.pewinternet.org/dataset/march-2016-libraries/}} and the survey of the Global Web Index (GWI)\footnote{\url{https://www.globalwebindex.net/}}. These datasets allow us to compute reference measurements of Facebook penetration and FGD for small samples of countries, to be compared with our calculation of the FGD through the Facebook marketing API.

The results of this validation exercise are reported in detail in Supplementary Text 1. We find high correlation coefficients between our measurement of penetration and the equivalents in GWI and the Pew Global survey, and for the case of FGD as well. These correlations are as good as the correlations between survey datasets, showing that the Facebook API data has comparable quality but a much higher coverage in terms of countries and better temporal resolution.
We find low and non-significant correlations between the absolute difference between our measurement of FGD and the one from surveys, but nevertheless add a control for Facebook penetration in our models to make sure that our results are not an artifact of correlated errors in the quantification of FGD.

We further compare Facebook penetration across ages groups in the US through the Pew Internet \& Technology survey and the GWI. We find very high correlations between age-dependent measurements. In addition, we explore how representative is the FGD for gender divides in other social media, as captured by the GWI survey. We found moderate yet significant correlations with other media such as WhatsApp, Twitter, and YouTube. This shows that, while we should not take Facebook as representative for all social media, there is certain similarity in gender differences that can motivate future research.

Finally, we test for intra-day oscillations of the measurement of FGD and Facebook penetration and found extremely consistent values. For the case of the FGD and the network externalities model, we also repeat our analysis on 
monthly snapshots of Facebook data for a period of twelve months between 2015 and 2016, calculating median DAU values each month. This way, we can confirm the robustness of our analysis to possible temporal changes in the way Facebook reports data through their API.

\subsection*{Gender equality and development datasets}

To normalize the number of active users over the total population of each country, we use the data collected by the US Census Bureau International Database\footnote{\url{https://www.census.gov/programs-surveys/international-programs/about/idb.html}}.
This dataset contains estimates of the resident population by age and gender for more than 226 countries.
We combine this data with gender equality indices measured by the World Economic Forum Gender Gap reports of 2015 and 2016 \cite{WEF}.
This dataset quantifies the magnitude of gender equality in 145 countries, measuring it with respect to four key areas: health, education, economic opportunity, and politics.
This report updates the values for education, economic, and political gender equality on a yearly basis, allowing us to measure changes between 2015 and 2016.
To account for additional economic and development indicators, we include data from the World Bank and the Human Development Index \cite{HDI}, measuring control variables of GDP (PPP) per capita in 2012, economic inequality as the quintile ratio, and Internet penetration.

\subsection*{Computing the Facebook Gender Divide}

We quantify the Facebook Gender Divide as a comparison of the rates of activity between genders.
The DAU measures how many users have logged into Facebook at a given day, which could be either through a Web browser or a mobile application.
We use the segmented data from 13 to 65 years old to normalize the DAU over the total population of a country in those ages, truncating all data that is not included in that age range.
This way we avoid introducing a bias with life expectancy and average age.
To have a stable estimate of the DAU, we use as the median value over the month of July 2015, replicating over other months afterwards.
This way, for each country $c$ and gender $g \in [Female,Male]$\footnote{For simplicity, we take gender as birth sex, i. e. male or female.}, we have a measurement of the number of active users $A_{g,c}$ between 13 and 65 years old. Additionally, this allows us to calculate the mean user age for a country to include it in our models.

Using the US Census Bureau data, we calculate the total population of each gender between the ages of 13 and 65 years old in each country, which we denote as $P_{g,c}$. This way, we can normalize the total activity in Facebook over the population in the same age ranges, calculating the \emph{activity ratios} $R_{g,c} = A_{g,c} / P_{g,c}$.
We define the \textbf{Facebook Gender Divide} in country $c$ as 
\begin{equation*}
FGD_c = log \left ( \frac{R_{Male,c}}{R_{Female,c}} \right )
\end{equation*}
, which compares male and female Facebook activity rates over the population of country $c$.
A country with positive FGD will have a tendency for men to be more present on Facebook, while a country with negative FGD will show the opposite tendency. A country with $FGD=0$ will have complete equality in the activity tendencies of both genders.

We further compute the \emph{Facebook penetration} as the ratio between user accounts between 13 and 65 years old reported by the API (regardless of activity and gender) and the total population of the country between those ages.

\subsection*{Regression Models}

We model dependencies between gender equality indicators and the FGD as linear models, after applying a rank transformation to all variables such that rank 1 is the highest possible value of the variable.
This way, we explore monotonic dependencies that do not need to be linear.
We define this \textit{FGD model} as:
\begin{equation*}
FGD = a_f \cdot Q + b_f \cdot C + c_f + \epsilon   \label{eq:FGD}
\end{equation*}
where $Q$ is a matrix with the ranks of economic, health, education, and political gender equality in each country and $C$ contains control variables such as Internet penetration (IP), income inequality (Ineq), total population (Pop), Facebook penetration (FBP) and mean user age (Age). $c_f$ is the intercept and $\epsilon$ denotes the residuals as the normally distributed, uncorrelated error of the model.

We analyze the relationship between changes and levels in economic gender equality and of FGD through with two models. First, an \textit{equality changes model}:
\begin{equation*}
\Delta Eco_{2016} = a_o \cdot Eco_{2015} + b_o \cdot FGD_{2015} + c_o \cdot O + d_o + \psi_o \label{eq:changes}
\end{equation*}

And second, a \textit{FGD changes model}:
\begin{equation*}
\Delta FGD_{2016} = a_q \cdot FGD_{2015} + b_q \cdot Eco_{2015} + c_q \cdot O + d_q + \psi_q \label{eq:changes2}
\end{equation*}

where $\Delta Eco_{2016}$ and $\Delta FGD_{2016}$ is the change in economic gender inequality and FGD between 2015 and 2016.
Both models include a control for autocorrelation as a term with the unranked value of the variable in the 2015, and a main term of the rescaled ranked value of the other variable. 
Following previous Economics research on Facebook data \cite{Gee2017}, we include various ranked controls in the matrix $O$, first with a simple correction for GDP, but then with extensions with other controls as for the FGD model.

We report the coefficient estimates of robust MM regressors for both models. To compare the effects of one variable over the changes in the other, we first residualize the changes by fitting against all controls. Then we compute the partial $R^2$ value of the conditioning variable when fitting the residualized values. To understand the uncertainty of this analysis, we bootstrap over $10.000$ samples and report the distribution of $R^2$ values.

We model network externalities as a power-law relationship between the activity ratio of a gender ($R_{g,c}$) and the total Facebook penetration for both genders together ($P_{c}$) in a joint model that includes an intercept for gender and interaction with gender. We define this way the \textit{network externalities model} as:

\begin{equation*}
 log(R_{g,c})  =  \alpha \cdot log(P_c) + \beta + \delta_{g,Female} (\alpha_F \cdot log(P_c) + \beta_F) +  \phi  \label{eq:network}
\end{equation*}

where $\alpha$ measures the scaling relationship between the Facebook presence ratio and the activity ratio of male users, $\alpha_F$ the difference in that relationship for female users, and $\phi$ the residuals. The Kronecker delta function  $\delta_{g,Female}$ takes value $1$ when $g=Female$ and $0$ otherwise.

All the above models do not show relevant multicollinearity when measuring Variance Inflation Factors \cite{Chatterjee2015}.

We report the fit the FGD model and the network externalities model with Markov Chain Monte Carlo sampling in JAGS \cite{Plummer2003}. 
We also fit all models with robust regression \cite{Koller2011}, reporting the results of the changes models in the main text and of the rest in the Supplementary Information.

To test the validity of the assumptions of our models after fitting, we verify the normality of residuals through Shapiro-Wilk tests \cite{Cromwell1994}, and check that residuals are uncorrelated with fitted values and independent variables.
For the case of the network externalities model, we additionally analyze multiplicative residuals in order to test for the possible role of outliers, as shown more in detail in Supplementary Information.

\section*{Acknowledments}

D.G. acknowledges funding from the Vienna Science and Technology Fund through the Vienna Research Group Grant ``Emotional Well-Being in the Digital Society'' (VRG16-005). E.M. acknowledges funding from Ministerio de Econom\'{\i}a y Competividad (Spain) through projects FIS2013-47532-C3-3-P and FIS2016-78904-C3-3-P. A.C. acknowledges funding from the H2020 EU project TYPES  (grant no. 653449) and the Ram\'{o}n y Cajal grant (RyC-2015-17732). Y.M.K. acknowledges funding from the H2020 EU project TYPES  (grant no. 653449). R.C. acknowledges funding from H2020 EU project ReCRED (grant no. 653417). We thank the Pew Research Center and the Global Web Index for the access to their data to validate the FGD.

\newpage

\section*{Supplementary Text 1 - Validating the FGD}

We validate the measurement of the FGD against three reference datasets:

\begin{enumerate}
	\item \textbf{Global Web Index (GWI).} We use the survey responses of the Global Web Index \footnote{\url{https://www.globalwebindex.net/}} panel during the period of our study (the two last quarters of 2015 and the two first quarters of 2016). For this period, the GWI contains responses from 99,338 panelists in 34 countries, providing rescaled estimates of survey responses that generalize to the population as a whole. We take the response to the question about the frequency of use of Facebook, calculating the weighted fraction of respondents for each gender and age segment that report to use Facebook daily or more than once a day. Furthermore, we repeat the analysis for other social media (Whatsapp, LinkedIn, Twitter, Instagram, and YouTube), calculating gender divide values outside Facebook.

	\item \textbf{Pew Research Center Spring 2016 Global Attitudes (Pew Global).} This dataset includes responses from 23,462 panelists in 19 countries abd can be found online \footnote{\url{http://www.pewglobal.org/dataset/spring-2016-survey-data/}}. We use the positive answer rate to question 82  ``Do you ever use online social networking sites like Facebook, Twitter...?'' as way to measure the penetration of SNS in general. We take the respondent weights reported in the dataset to rescale the frequency of positive responses taking into account the self-reported gender of survey respondents.

	\item \textbf{Pew Research Center Internet \& Technology, March 7-April 4, 2016 (Pew US).} This US questionnaire \footnote{\url{http://www.pewinternet.org/dataset/march-2016-libraries/}} includes questions about Facebook use in particular (act135,  ``Do you ever use the internet or a mobile app to use Facebook?''), as well as gender and age data from 1,601 respondents. We use the respondent weights to compute Facebook use rates across gender and age groups.

\end{enumerate}

\subsection*{Comparison of Facebook penetration estimates}

\begin{figure}[h]
\centering
\includegraphics[width=0.95\textwidth]{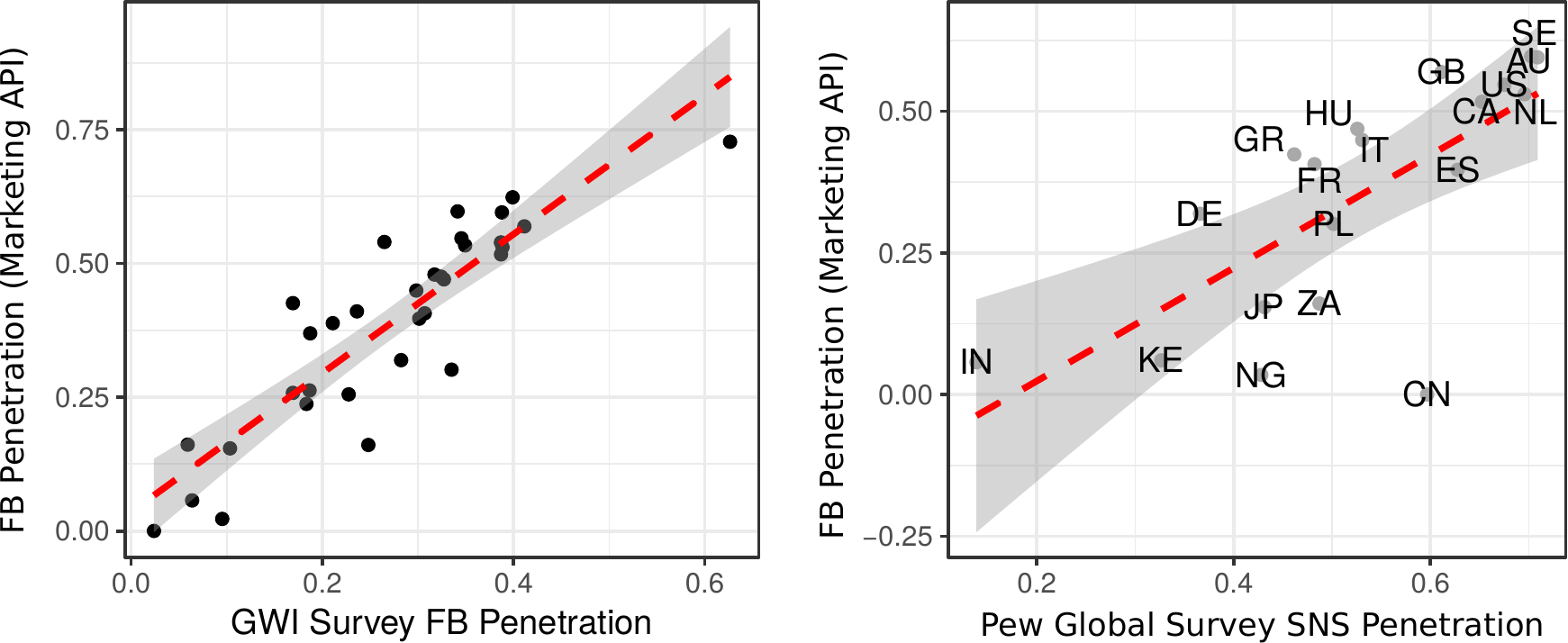}
\caption{Total Facebook penetration for both genders as reported by the marketing API versus Facebook penetration as estimated in the GWI survey (left) and penetration of all Social Networking Sites in Pew Global survey (right). The red dashed line shows a linear regression profile, with its prediction standard errors in the shaded area.
\label{fig:PenetrationValidation}} 
\end{figure}

The left panel of Fig. \ref{fig:PenetrationValidation} shows the relationship between the total Facebook penetration for both genders as measured by us through the marketing API versus the value estimated from the GWI survey.
There is a high positive correlation between both measurements of Facebook Penetration (Pearson $0.89$ , CI $[0.78,0.94]$, Spearman $0.86$, CI $[0.72,0.94]$).
The right panel of Fig. \ref{fig:PenetrationValidation} shows the same evaluation against the Pew Global survey. The correlation between both measurements is positive and high (Pearson: $0.71$, CI $[0.39,0.88]$, Spearman: $0.77$, CI $ [0.43,0.94]$).

\begin{figure}[h]
\centering
\includegraphics[width=0.95\textwidth]{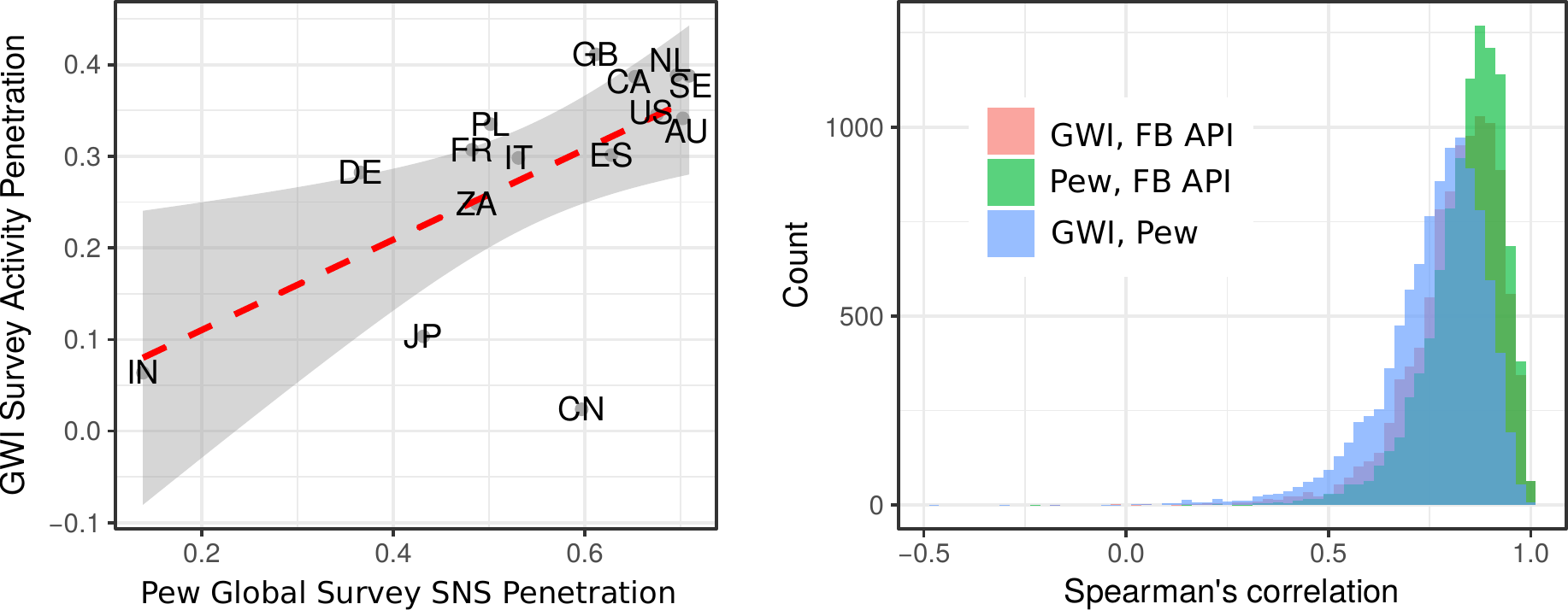}
\caption{Left: Comparison of GWI Facebook penetration estimate and Pew penetration of all SNS. The red dashed line shows a linear regression profile, with its prediction standard errors in the shaded area. Right: Bootstrapping distributions of Spearman's correlation coefficient between all pairs of penetration measurements. 
\label{fig:PenetrationValidation2}} 
\end{figure}

The left panel of Fig. \ref{fig:PenetrationValidation2} shows the comparison between estimates based on GWI and Pew Global survey data. The correlation is also positive and significant, even though samples are of limited size (Pearson: $0.63$, CI $[0.17,0.86]$, Spearman: $0.7$, CI $[0.38,0.91]$)
Both in the right panel of Fig. \ref{fig:PenetrationValidation} and in left panel of Fig. \ref{fig:PenetrationValidation2}, it can be seen that China is a clear outlier. This stems from the difference between comparing Facebook penetration versus penetration for Social Networking Sites in general, which is the precise question of the Pew survey. If we focus on the rest of countries, where we can expect a priori that Facebook is more representative of social media in general, the correlation reaches higher values when comparing to the Facebook marketing API (Pearson: $0.84$, CI $[0.62,0.94]$, Spearman: $0.87$, CI $[0.66,0.96]$) and to the GWI survey data (Pearson $0.84$, CI $[0.55,0.95]$, Spearman $0.78$, CI $[0.43,0.94]$). To make a fair comparison, we should not include China when comparing Facebook penetration with penetration in SNS in general.

When we focus only on the set of countries available in all three datasets, we can compare the correlation of each pair of data sources to assess the validity of the Facebook API data in terms of Facebook penetration. In this smaller sample, the Facebook penetration calculated through the marketing API is highly correlated with both
the Pew Global survey values (Pearson: $0.87$, CI $[0.64,0.96]$, Spearman: $0.86$, CI $[0.59,0.97]$) and with the GWI survey (Pearson: $0.89$, CI $[0.68,0.96]$, Spearman:  $0.83$, CI $[0.53,0.97]$). The point estimates of these two values are higher than the correlation between the Pew Global and GWI datasets, as reported above. Nevertheless, these differences are not significant, as confidence intervals overlap and bootstrap sampling shows that estimates are indistinguishable (Fig. \ref{fig:PenetrationValidation}). From this analysis we conclude that the estimate of Facebook penetration from the marketing API has comparable quality to the values reported in the high-quality, representative surveys of GWI and the Pew Research Center.

\subsection*{Facebook gender divide estimates}

\begin{figure}[h]
\centering
\includegraphics[width=0.95\textwidth]{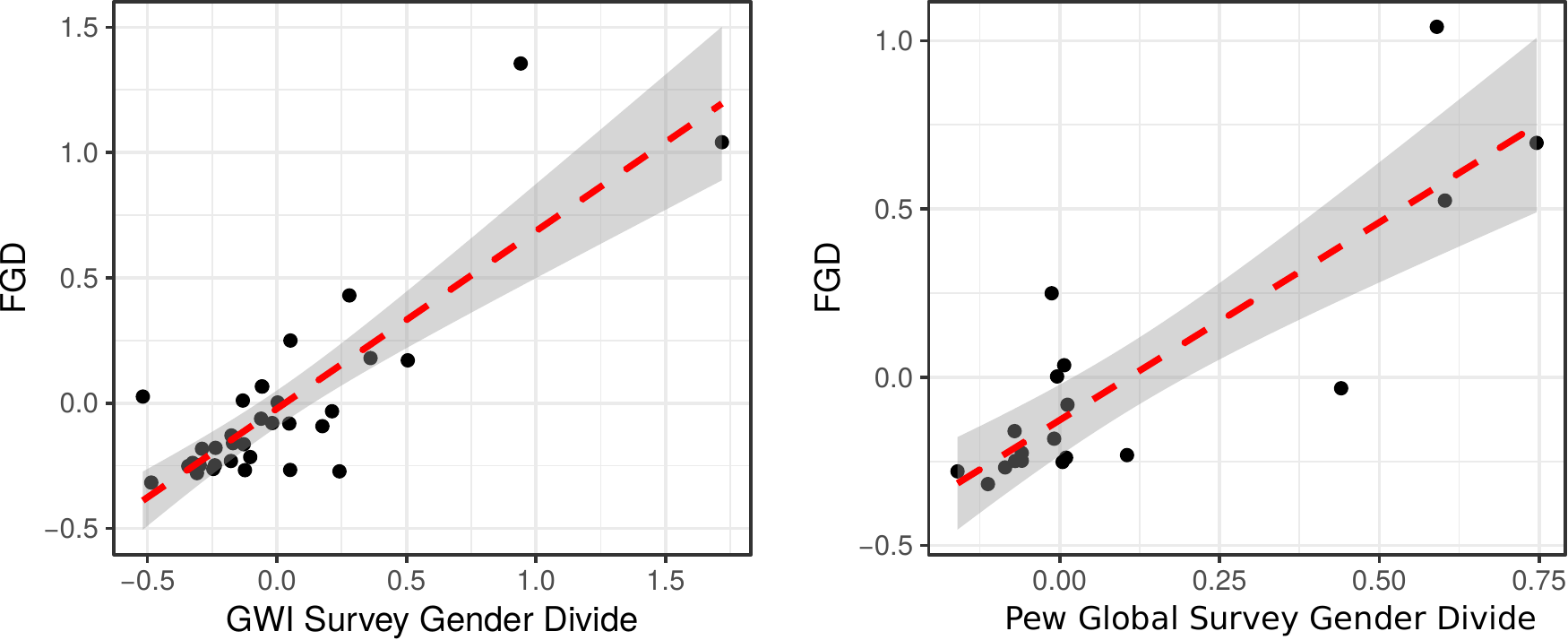}
\caption{Measurement of the FGD in the API versus estimates using GWI data (left) and a gender divide for all SNS in PEW (right).
The red dashed line shows a linear regression profile, with its prediction standard errors in the shaded area.
\label{fig:FGDValidation}} 
\end{figure}

We calculated surrogates of the FGD based on the GWI and Pew global survey datasets. The left panel of Fig. 
\ref{fig:FGDValidation} shows the comparison of the Facebook marketing API estimate with the same estimate using GWI data, revealing high positive correlation (Pearson: $0.83$, CI $[0.68,0.91]$, Spearman: $0.63$, CI $[0.27,0.87]$). The right panel of Fig. \ref{fig:FGDValidation} shows the comparison of the Facebook marketing API estimate with Pew survey data for all SNS (including China), also revealing high positive correlation (Pearson: $0.85$, CI $[0.65,0.94]$, Spearman:  $0.74$, CI $[0.35,0.91]$).

\begin{figure}[h]
\centering
\includegraphics[width=0.95\textwidth]{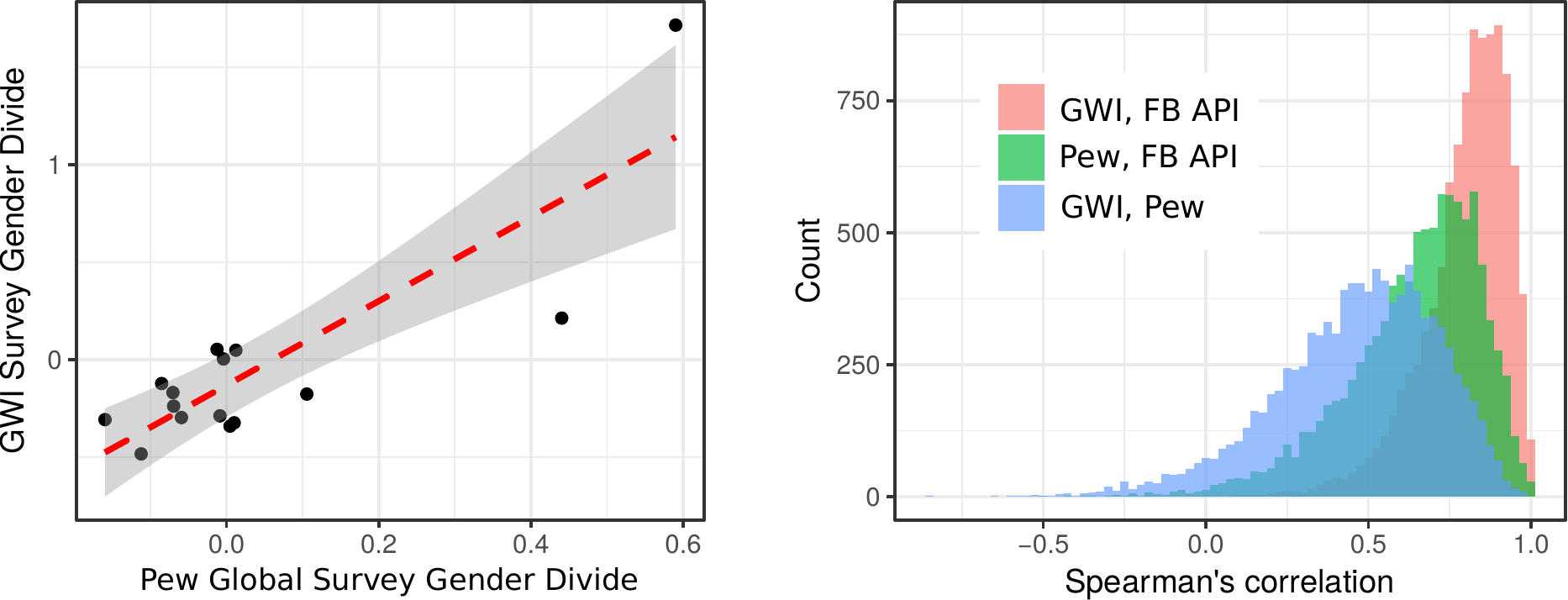}
\caption{Left: Comparison of gender divide measurements in PEW and GWI data. The red dashed line shows a linear regression profile, with its prediction standard errors in the shaded area. Right: Bootstrapping distributions of Spearman's correlation coefficient between all pairs of gender divide measurements. \label{fig:FGDValidation2}} 
\end{figure}

A comparison between estimates of the FGD using GWI versus using Pew data is shown on the left panel of Fig.  \ref{fig:FGDValidation2}, also revealing positive correlations (Pearson: $0.85$, CI $[0.6,0.95]$, Spearman: $0.49$, CI $[-0.11,0.86]$).
As with Facebook Penetration, the correlation between the FGD using the Facebook marketing API and the other two (with Pew: Pearson: $0.77$, CI $[0.43,0.92]$, Spearman: $0.68$, CI $[0.14,0.93]$; with GWI: Pearson  $0.96$, CI $[0.89 0.99]$, Spearman $0.83$, CI $[0.47,0.97]$) estimates is comparable to the correlation within estimates, as evidenced in bootstrapping samples reported in the right panel of Fig. \ref{fig:FGDValidation2}. We can conclude that the estimate of the FGD using the marketing API is consistent with GWI and Pew survey metrics, opening the study of the FGD to a much larger sample of countries.

We measured the absolute difference between the FGD in the marketing API and in each survey dataset. As expected, the correlation between this absolute difference and the Facebook penetration across countries is negative (with GWI: Pearson: $-0.3$, CI $[-0.59,0.05]$, p-value$= 0.09$; with Pew: Pearson: $-0.27$, CI $[-0.65,0.21]$, p-value$= 0.26$), but its value is weak and not significant. Nevertheless, we include controls for Facebook penetration in our further analyses, to make sure that our results are not an artifact of a correlation between penetration and measurement error in the marketing API.

Furthermore, the GWI survey allows us to compare measurements of the FGD in other social networks with our measure based on the Facebook marketing API. We get moderate to high Pearson correlation coefficients with other sites, such as Whatsapp ($0.67$, CI $[0.43,0.82]$), LinkedIn ($0.65$, CI $[0.40,0.81]$), Twitter ($0.69$, CI $[0.46,0.84]$), Instagram  ($0.79$, CI $[0.62,0.89]$), and YouTube ($0.89$, CI $[0.79,0.94]$). While we cannot generalize to all social networks based only on Facebook data, we can see that, to some extent, the difference in activity across genders also appears in other SNS. This is particularly interesting when comparing Facebook, a very private social network, with YouTube or Twitter, which are much more public but still display substantial correlations in terms of FGD.

\subsection*{Comparison across age groups}

\begin{figure}[h]
\centering
\includegraphics[width=0.95\textwidth]{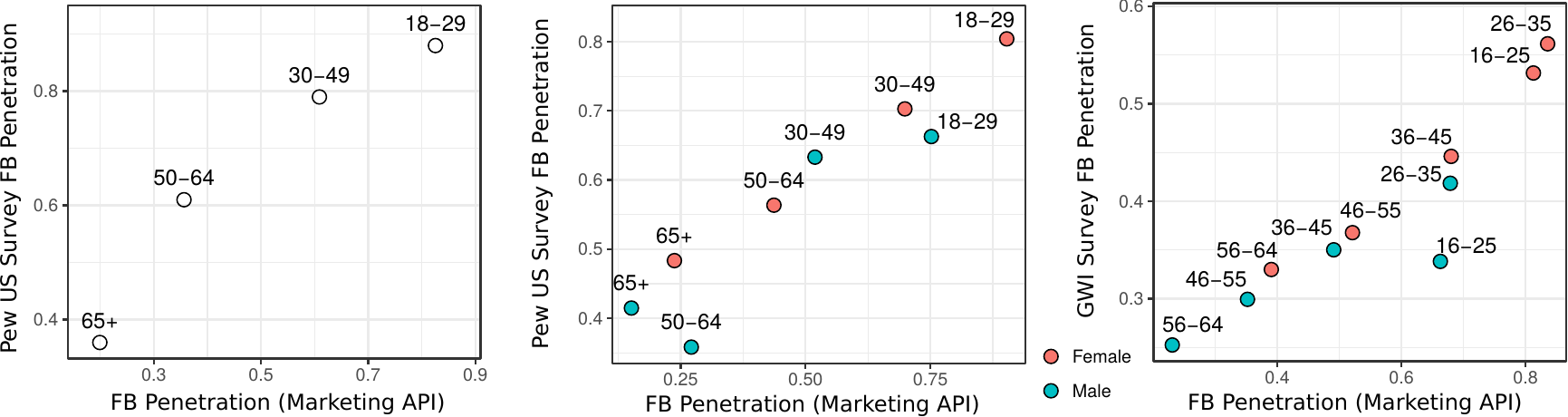}
\caption{Comparison of FB presence ratio versus PEW gender categories for both genders together (left) and gender-wise (center). Replication of the same validation versus GWI estimates (right).\label{fig:AgeValidation}} 
\end{figure}

We compared Facebook penetration estimates in the US across the four age groups reported in the Pew US dataset. The left panel of Fig. \ref{fig:AgeValidation} shows the comparison for both genders together, which have a Pearson correlation coefficient of $0.96$, CI $[0.04,0.99]$. The central panel of Fig. \ref{fig:AgeValidation} shows the same comparison by taking the gender-wise estimates, which also have positive Pearson correlation ($0.94$, CI $[0.72,0.99]$).
This also appears when surveying the GWI dataset for US respondents in similar age categories, as shown on the left panel of Fig. \ref{fig:AgeValidation}, which has a high and significant Pearson correlation coefficient of $0.92$, CI $[0.69,0.98]$. We can conclude that data provided by the Facebook marketing API is consistent across ages, but to be sure that our further analyses are robust we take two action: 1) we add a mean user age control to our regression models, and 2) we stratify our analyses across age categories, using in each stratum a measurement of the FGD in the corresponding age range.

\subsection*{Subdaily measurement consistency}

\begin{figure}[h]
\centering
\includegraphics[width=0.95\textwidth]{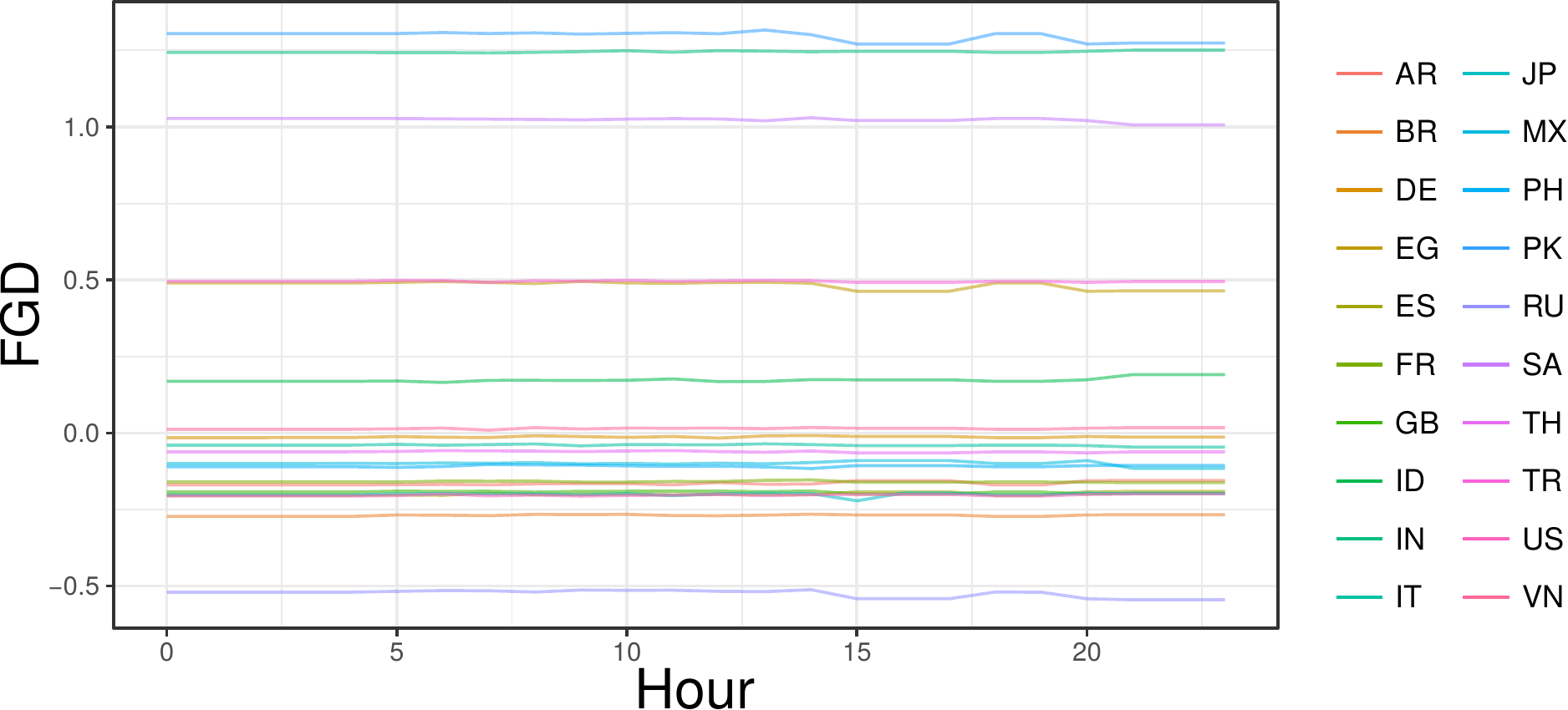}
\caption{Examples of subdaily trajectories of the FGD. \label{fig:HourValidation1}} 
\end{figure}

We retrieved data from the Facebook marketing API on a daily frequency, starting our retrieval at 3 AM Central European Time. To validate the consistency of our measurement with any other times of the day, we checked the consistency of our construction of the FGD with hourly values for 24 hours in December 2017. 

\begin{figure}[h]
\centering
\includegraphics[width=0.95\textwidth]{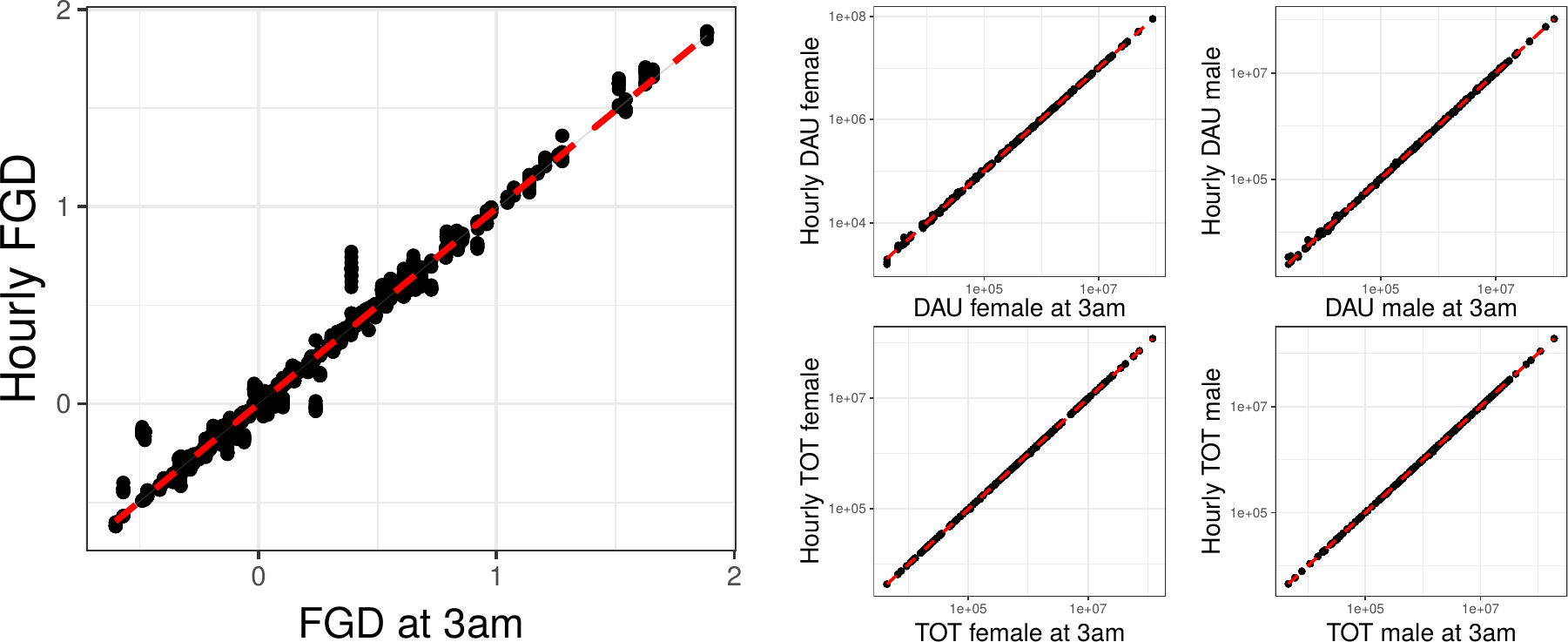}
\caption{Facebook API measurements at different hours of the day. The left panel shows a comparison of measurements of the FGD for all countries in the dataset at 3AM CET versus hourly measurements at other times of the day. The right panels show the comparison between the measurement at 3AM CET and at other times of the day for the number of DAU and of present users (TOT) per gender. \label{fig:HourValidation2}} 
\end{figure}

Fig. \ref{fig:HourValidation1} shows the hourly measurement for a sample of large countries, revealing high consistency with very small fluctuations. When comparing the measurement of the FGD at 3AM CET with any other time in the same day, we get extremely high pearson correlation coefficients ($0.9942654$, CI $[0.9939277,0.9945844]$), as also evidenced in Fig. \ref{fig:HourValidation2}. This also extends to the measurement of Daily Active Users for male ($0.9999431$, CI $[0.9999397,0.9999462]$) and female ($0.9999378$, CI $[0.9999341,0.9999412]$), as well as the total number of accounts for male users ($0.9999926$, CI $[0.9999921,0.9999930]$) and female users ($0.9999968$, CI $[0.9999966,0.9999970]$). Any fluctuation can be attributed to the rounding that Facebook does to preserve individual user anonymity and to the inter day changes in the number of Daily Active Users

\newpage

\section*{Supplementary Text 2 - FGD as a function of other inequalities}

\subsection*{Regression diagnostics}

The Variance Inflation Factors of the variables in the FGD model are below 5, allowing us to discard collinearity in the linear model of FGD as a function of other inequalities.
Table \ref{tab:FGDmodel} reports the detailed results of the FGD model fit and Table \ref{tab:FGDmodelRobust} reports the results of the same model when fitted with a robust regression method. Table \ref{tab:FGDmodelCor} shows a fit with HC correction for heteroskedasticity. All results are qualitatively similar, revealing that the FGD model result is robust to outliers and heteroskedasticity.

\begin{table}[h!]
\centering
\begin{tabular}{|c c c c|}
\hline
Term   &   Median estimate   & 95\% Credible Interval  & p-value \\
\hline
Intercept & $\bf  135.8$ & $[119.9,152.1]$ & $p<0.01$ \\
Education Equality Rank & $\bf -0.54$ & $[-0.67,-0.41]$ & $p<0.01$ \\
Health Equality Rank & $\bf -0.27$ & $[-0.37,-0.17]$ & $p<0.01$ \\
Economic Equality Rank & $\bf -0.16$ & $[-0.27,-0.06]$ & $p<0.01$ \\
Political Equality Rank & $0.05$ & $[-0.05,0.14]$ & $0.19$ \\
Internet Penetration Rank & $\bf -0.27$ & $[-0.44,-0.09]$ & $p<0.01$ \\
Income Inequality Rank & $0.01$ & $[-0.09,0.10]$ & $0.44$ \\
Population Rank & $0.01$ & $[-0.09,0.11]$ & $0.40$ \\
Facebook Penetration Rank & $0.03$ & $[-0.12,0.18]$ & $0.33$ \\
Mean User Age Rank & $0.02$ & $[-0.08,0.11]$ & $0.33$ \\
\hline\hline
$N$ & 142 & $R^2$ & $ 0.7417$ \\
\hline
\end{tabular}
\caption{Regression results of FGD model. Estimates of p-values are based on the posterior of parameter estimates after 10,000 iterations. \label{tab:FGDmodel}}
\end{table}

\begin{figure}[h!]
\centering
\includegraphics[width=0.66\textwidth]{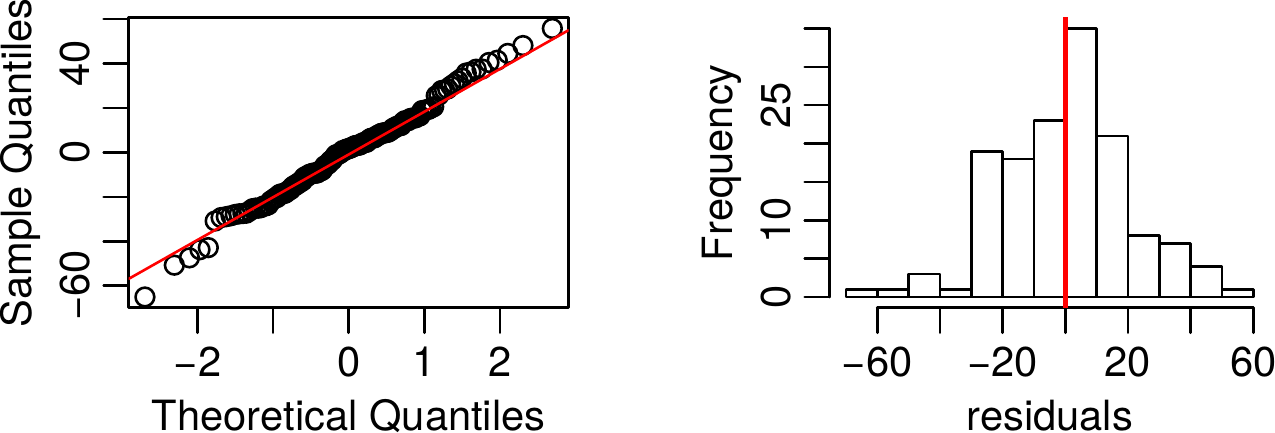}
\caption{Left: Normal Q-Q plot of residuals of the FGD model. Right: Histogram of residuals.\label{fig:FGDResiduals}} 
\end{figure}

Fig. \ref{fig:FGDResiduals} shows the normal Q-Q plot and the histogram of residuals, which are distributed very close to normality. This is confirmed by a Shapiro-Wilk normality test, with a statistic of 0.99 and unable to reject the null hypothesis that residuals are normally distributed ($p=0.63$).
Furthermore, residuals are uncorrelated with all gender equality variables (Near-zero Pearson correlation coefficients, with p-values above $0.9$) and the square root of absolute residuals are not significantly correlated with predicted values.
In addition, Facebook penetration is uncorrelated with residuals of the model (Pearson $-0.0028$, p-value $=0.97$), showing no signs of bias due to the variance of Facebook penetration rates across countries.

\begin{table}[h!]
\centering
\begin{tabular}{|c c c c c|}
\hline
Term   &   Estimate   & Standard Error & t-value & p-value \\
\hline    
Intercept        &  $\bf 137.4$   & $9.14$ & $15.04$ & $p< 10^{-10}$ \\
Education Equality Rank  &  $\bf -0.55$ &  $0.07$ & $-7.47$ & $p< 10^{-10}$ \\
Health Equality Rank  &    $\bf -0.29$ & $0.05$ & $-4.98$ & $p< 10^{-5}$ \\ 
Economic Equality Rank  &  $\bf -0.13$ &  $0.06$ & $-2.03$ &  $0.04$  \\ 
Political Equality Rank  &  $0.05$  & $0.05$  & $0.95$ &  $0.34$    \\
Internet Penetration Rank & $\bf -0.29$  & $0.10$ & $-2.89$ & $p< 0.01$ \\
Income Inequality Rank  &      $0.001$ &  $0.05$ & $0.01$ &  $0.99$   \\ 
Population Rank &  $-0.003$ &  $0.05$ & $0.05$  & $0.96$    \\
Facebook Penetration Rank &  $0.05$ &  $0.09$ & $0.62$  & $0.54$    \\
Mean User Age Rank &  $0.01$ &  $0.05$ & $0.21$  & $0.84$    \\
\hline\hline
$N$ & 142 & \multicolumn{2}{c}{Multiple $R^2$} & $ 0.6715$  \\
\hline
\end{tabular}
\caption{Robust regression results of FGD model. \label{tab:FGDmodelRobust}}
\end{table}

\begin{table}[h!]
\centering
\begin{tabular}{|c c c c c|}
\hline
Term   &   Estimate   & 95\% HC CI  & Standard error & p-value \\
\hline
Intercept & $\bf  136.8$ & $[122.5,151.2]$ & $7.32$ & $p< 10^{-10}$ \\
Education Equality Rank & $\bf -0.54$ & $[-0.68,-0.40]$ & $0.07$  & $p< 10^{-10}$ \\
Health Equality Rank & $\bf -0.27$ & $[-0.36,-0.18]$ & $0.05$  & $p< 10^{-8}$ \\
Economic Equality Rank & $\bf -0.17$ & $[-0.26,-0.07]$  & $0.05$ & $p<0.001$ \\
Political Equality Rank & $0.04$ & $[-0.04,0.13]$ & $0.04$ & $0.32$ \\
Internet Penetration Rank & $\bf -0.27$ & $[-0.44,-0.10]$ & $0.09$ & $p<0.01$ \\
Income Inequality Rank & $0.004$ & $[-0.09,0.10]$ & $0.05$ & $0.92$ \\
Population Rank & $0.01$ & $[-0.08,0.10]$  & $0.05$ & $0.87$ \\
Facebook Penetration Rank & $0.03$ & $[-0.12,0.18]$ & $0.08$ & $0.7$ \\
Mean User Age Rank & $0.02$ & $[-0.07,0.10]$ & $0.04$  & $0.66$ \\
\hline
\end{tabular}
\caption{Coefficient estimates using HC corrected estimates. \label{tab:FGDmodelCor}}
\end{table}

\newpage

\subsection*{The role of GDP}

\begin{figure}[h]
\centering
\includegraphics[width=0.7\textwidth]{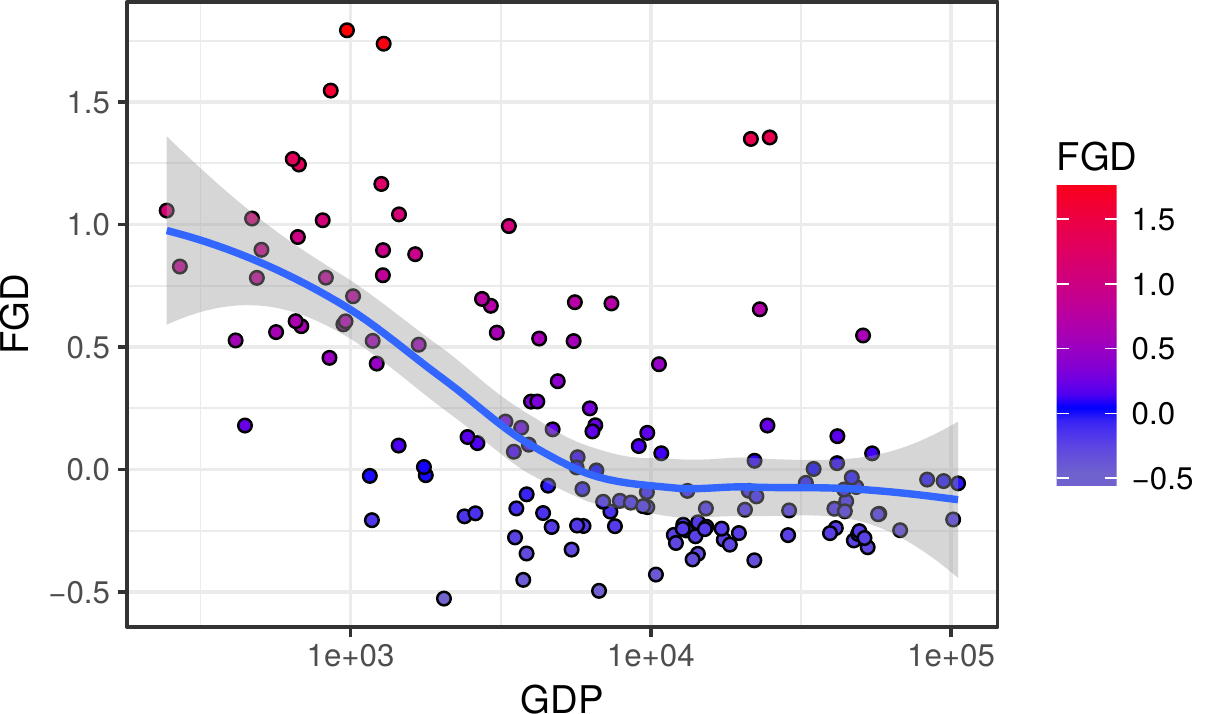}
\caption{Relationship between FGD and GDP (log scale). \label{fig:GDP-FGD}} 
\end{figure}

\begin{figure}[h]
\centering
\includegraphics[width=0.7\textwidth]{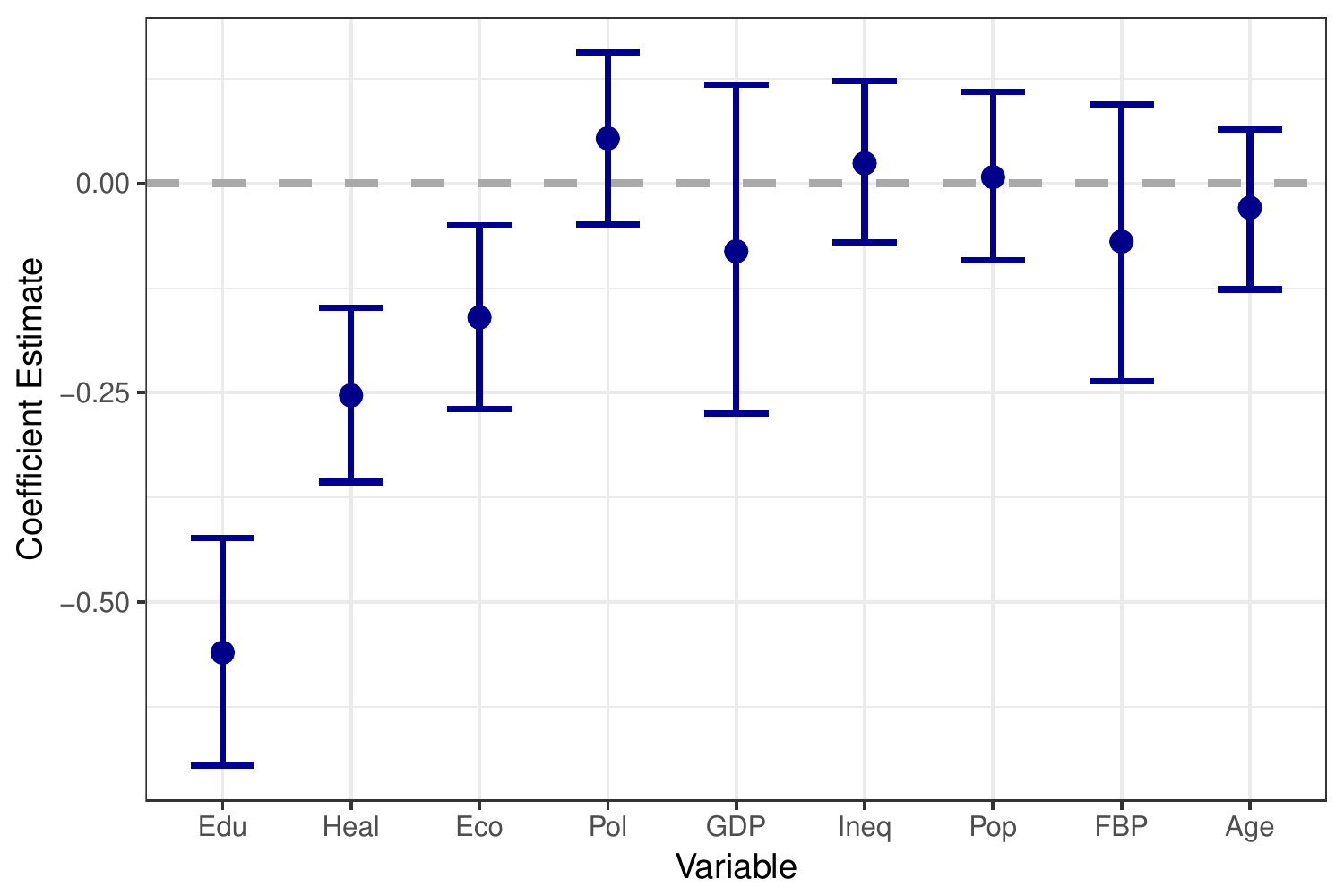}
\caption{Coefficient estimates of the FGD model with GDP instead of Internet Penetration as control. \label{fig:GDP-FGD-Model}} 
\end{figure}

Fig. shows the relationship between the FGD and the GDP per capita. There is a significant negative correlation between both (Spearman $-0.57$, $p < 10^{-6}$), motivating a replication of the above model with GDP as a control.
Since GDP is highly correlated with Internet Penetration and leads to high Variance Inflation, we replace Internet Penetration with GDP in our model. Coefficient estimates are reported on Fig. \ref{fig:GDP-FGD-Model}, revealing that the main result is robust to controlling for the wealth of countries.

\newpage
\subsection*{Model stability across monthly measurements}

\begin{figure}[h]
\centering
\includegraphics[width=0.8\textwidth]{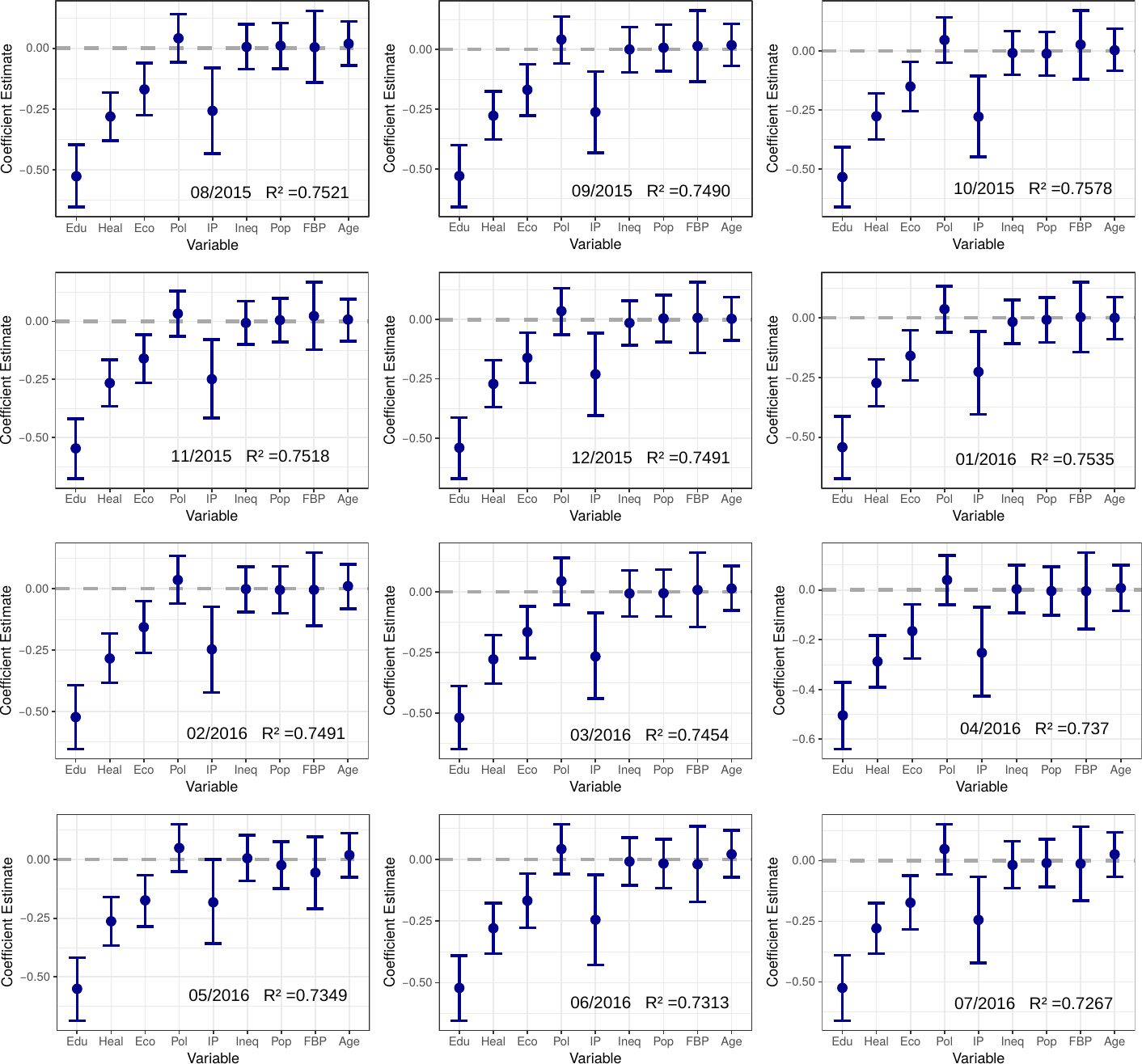}
\caption{Results of repetition of the fit of the FGD model for 12 months between 2015 and 2016. The results are qualitatively stable across months. \label{fig:FGDmodelMonth}} 
\end{figure}

We repeated the fit of the FGD model for measurements of the DAU in twelve months between 2015 and 2016. Fig. \ref{fig:FGDmodelMonth} shows the results of the fit for these alternative periods. The coefficient estimates barely depend on the period when the DAU are calculated and the $R^2$ of the fits range between $0.727$ and $0.758$, confirming that our results are robust to fluctuations in the reporting of DAU through the Facebook API.

\newpage

\newpage

\subsection*{Model stability across age groups}

\begin{figure}[h]
\centering
\includegraphics[width=0.85\textwidth]{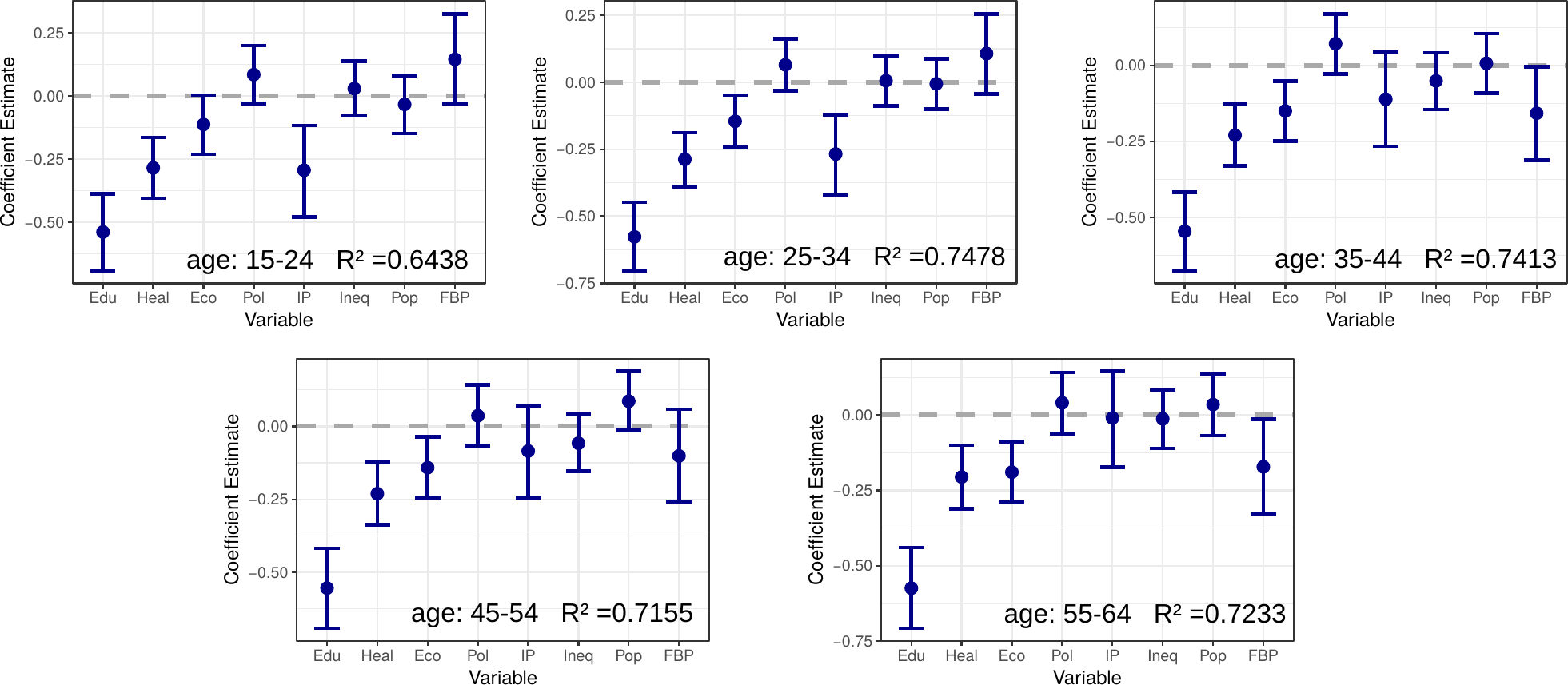}
\caption{Replication of the model for data segmented into age groups. \label{fig:FGDmodelAges}} 
\end{figure}

Figure \ref{fig:FGDmodelAges} shows the replication of the model for segments of different age groups. Results are qualitatively similar to those of the whole population, with significant effects of gender equality variables and high $R^2$ values. 

\newpage

\subsection*{Model test with GWI and Pew approximations to the FGD}

\begin{figure}[h]
\centering
\includegraphics[width=0.45\textwidth]{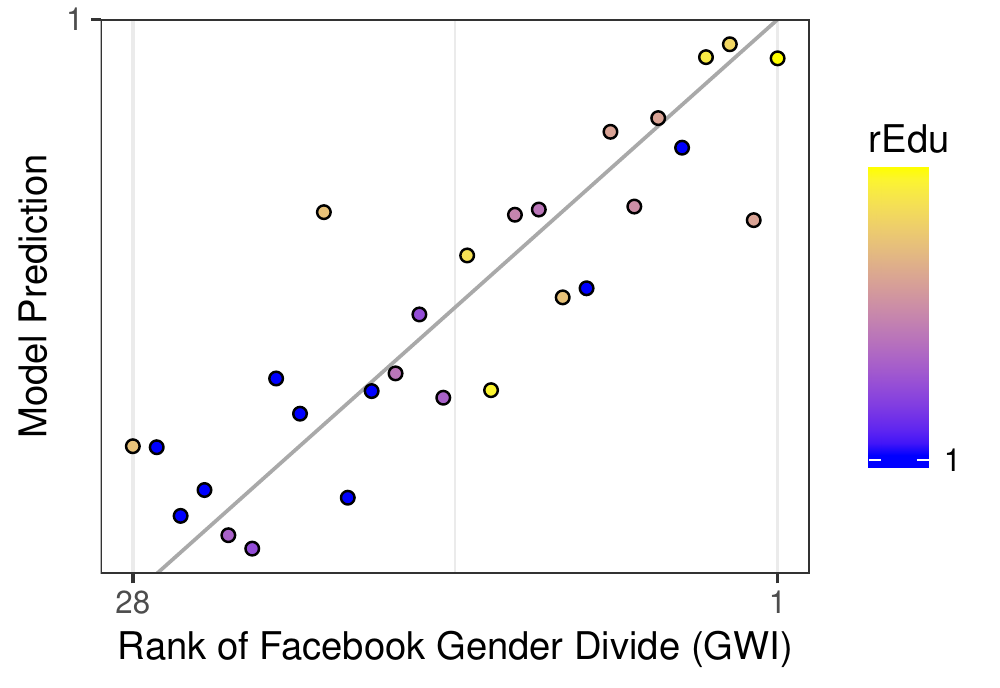}
\caption{Replication of the model with the GWI estimate of the FGD. \label{fig:GWIFGDmodel}} 
\end{figure}

We repeated the model using approximations of the FGD using the limited sample of GWI. The performance of the model is similar, as shown in Figure \ref{fig:GWIFGDmodel}, with $R^2=0.77$. While the sample size of GWI is too small to test the role of all equality variables, the Pearson correlation between the rank of education equality and the rank of FGD in GWI is $-0.52$ ($p-value<0.01$).
Similarly, the model for the approximation of the FGD with data from Pew for all SNS gives similar $R^2=0.63$ and a significant negative Pearson correlation between the rank of education equality and the rank of FGD in PEW ($-0.73$, $p-value<0.001$).

\newpage

\section*{Supplementary Text 3 - Network externalities}

\begin{figure}[h]
\centering
\includegraphics[width=1\textwidth]{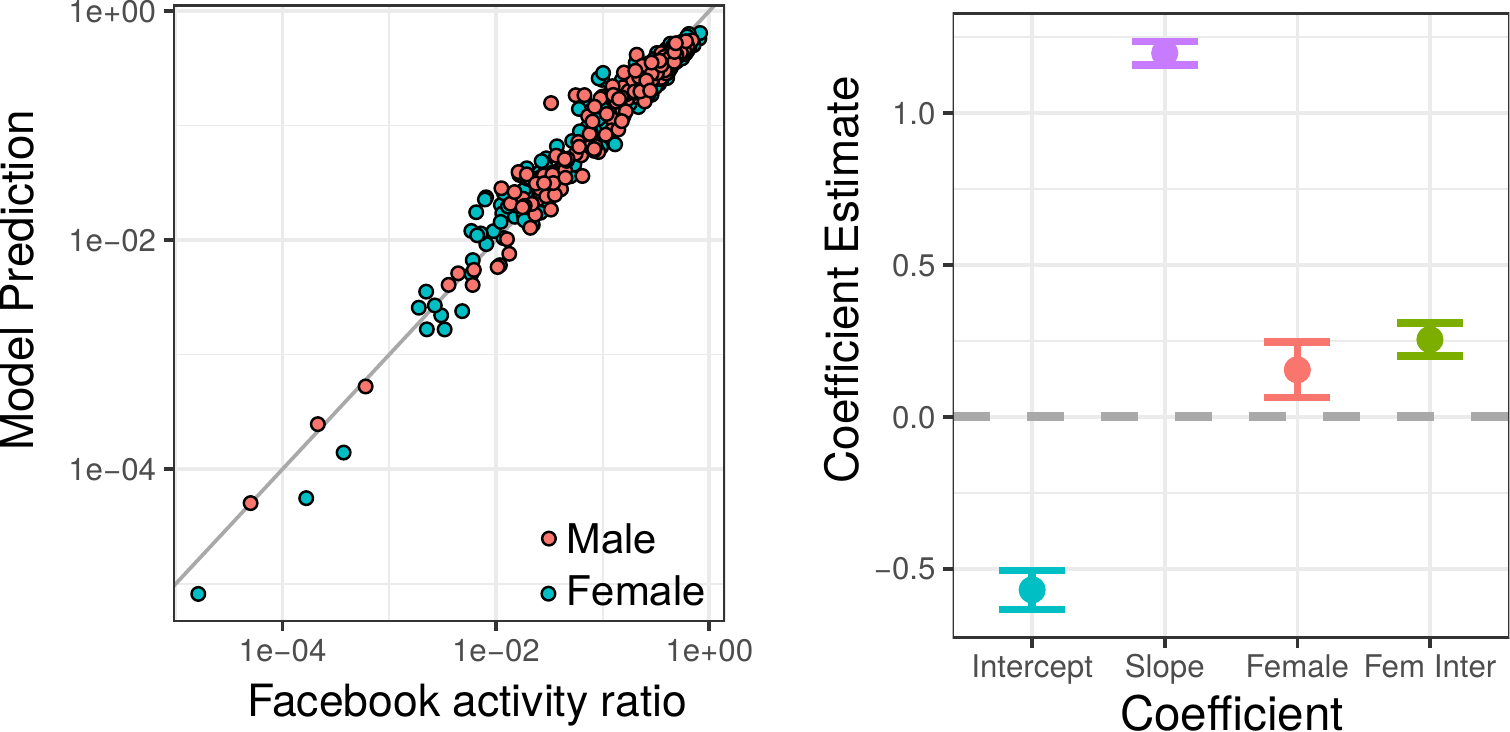}
\caption{Results of network externalities model fit. The left panel shows the comparison between empirical and predicted values, the right panel shows median estimates and 95\% CI of model coefficients. \label{fig:NetworkModel}} 
\end{figure}

The results of the network externalities model are shown on Fig. \ref{fig:NetworkModel}. The model achieves a $R^2=0.96$ on the logarithmic scale and a $R^2=0.89$ on the linear scale of activity ratios per gender. Table \ref{tab:networkmodel} shows the detailed results of the model, evidencing the superlinear scaling ($\alpha = 1.2$) and the difference between genders ($\alpha_F=0.25$).

\begin{table}[h]
\centering
\begin{tabular}{|c c c c|}
\hline
Term   &   Median estimate   & 95\% Credible Interval  & p-value \\
\hline
$\beta$ & $\bf -0.57$ & $[-0.63,-0.50]$ & $p<0.01$ \\
$\alpha$ & $\bf 1.198$ & $[1.16,1.24]$ & $p<0.01$ \\
$\beta_F$ & $\bf 0.15$ & $[0.06,0.25]$ & $p<0.01$ \\
$\alpha_F$ & $\bf 0.25$ & $[0.2,0.31]$ & $p<0.01$ \\
\hline\hline
$N$ & 422 (211 countries, 2 genders) & $R^2$ & $0.96$ \\
\hline
\end{tabular}
\caption{Regression results of network externalities model. Estimates of p-values are based on the null hypothesis that the coefficient equals one for $\alpha$ and zero for the rest, after 10,000 iterations. \label{tab:networkmodel}}
\end{table}

\begin{figure}[h]
\centering
\includegraphics[width=0.9\textwidth]{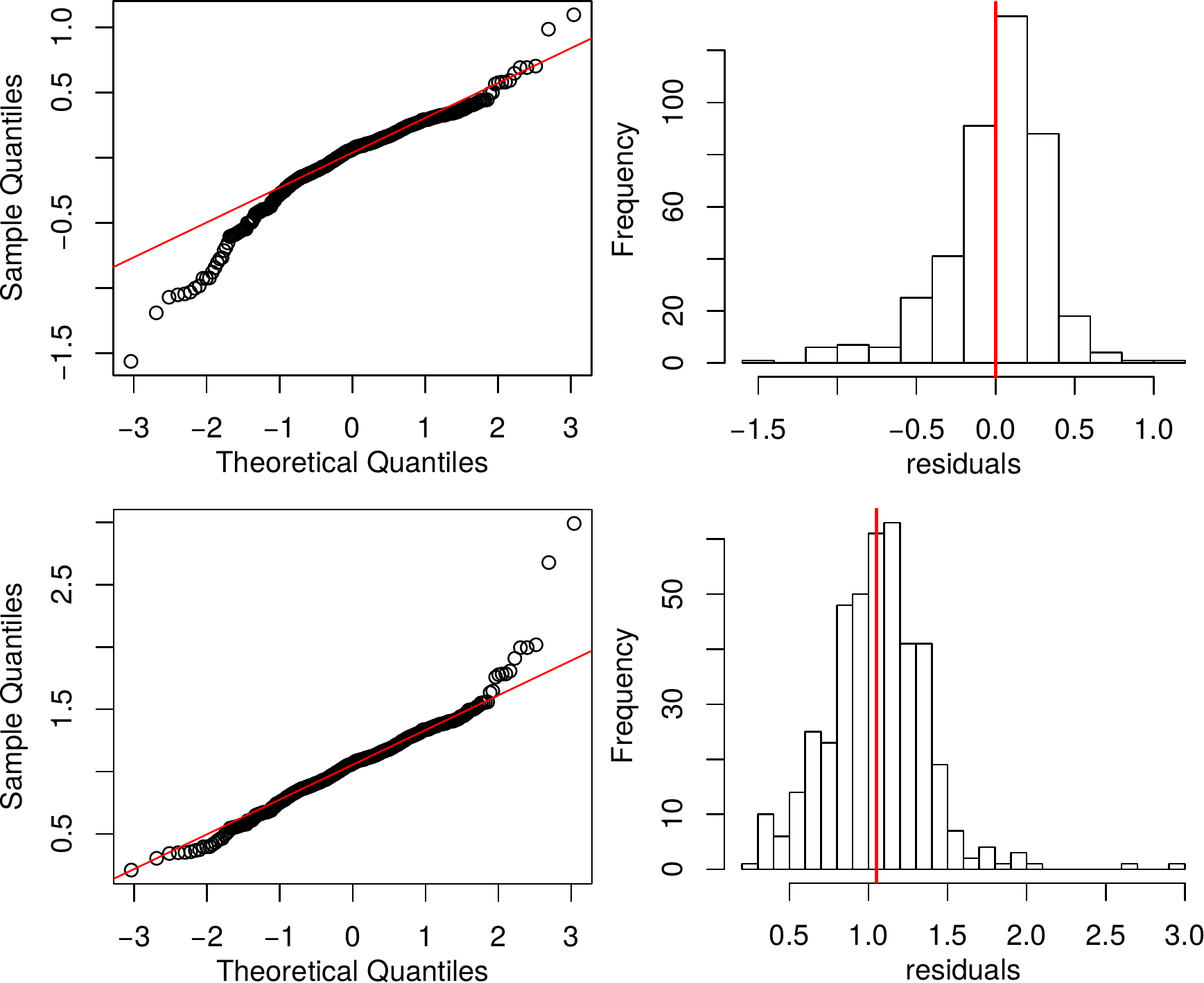}
\caption{Analysis of residuals of network externalities model. The top panels show the normal Q-Q plot and the histogram of residuals $\phi$ of the model, and the lower panels the converse for multiplicative residuals in the linear scale of activity ratios per gender. Some minor deviations from normality can be observed in both.
\label{fig:NetworkResiduals}} 
\end{figure}

Fig. \ref{fig:NetworkResiduals} shows the analysis of the residuals of the model and the error in the linear scale of activity ratios per gender. Some small deviations from normality can be observed at the tails, corresponding to significant Shapiro-Wilk statistics of 0.94 and 0.95.
Both types of residuals are uncorrelated with Facebook penetration and do not appear to have a structure across predicted values.
We identified some of the residual outliers, such as China and Tajikistan, which when removed do not have a qualitative impact in the results of the model fit and lead to residual distributions closer to normality.

Table \ref{tab:networkmodelHC} shows the results when correcting for heteroskedasticity. All results remain qualitatively unchanged.

\begin{table}[h]
\centering
\begin{tabular}{|c c c c c|}
\hline
Term   &   Estimate   & 95\% Confidence Interval & Standard Error & p-value \\
\hline
$\beta$ & $\bf -0.57$ & $[-0.61,-0.53]$ & $0.02$ & $p<10^{-10}$ \\
$\alpha$ & $\bf 1.198$ & $[1.17,1.23]$ & $0.02$ & $p<10^{-10}$ \\
$\beta_F$ & $\bf 0.15$ & $[0.07,0.24]$  & $0.045$ & $p<0.001$ \\
$\alpha_F$ & $\bf 0.25$ & $[0.18,0.33]$  & $0.038$ & $p<10^{-10}$ \\
\hline
\end{tabular}
\caption{HC corrected results of network externalities model. \label{tab:networkmodelHC}}
\end{table}

We repeated the fit using a robust regression method, reporting the results on Table \ref{tab:NetworkModelRobust}. While estimates slightly change, the qualitative results of a superlinear relationship that is stronger for female users still hold. This shows that our conclusions are robust to the influence of outliers.

\begin{table}[h]
\centering
\begin{tabular}{|c c c c c|}
\hline
Term   &   Estimate   & Standard Error & t-value & p-value \\
\hline    
$\beta$  &  $\bf -0.523$   & $0.033$ & $-16.25$ & $p< 10^{-10}$ \\
$\alpha$ &  $\bf 1.19$    & $0.02$ & $60.88$ & $p< 10^{-10}$ \\ 
$\beta_F$  &  $\bf 0.25$    & $0.05$ & $5.43$ & $p< 10^{-7}$ \\
$\alpha_F$ &  $\bf 0.37$    & $0.03$ & $12.59$ & $p< 10^{-10}$ \\
\hline
\end{tabular}
\caption{Robust regression results of the network externalities model. \label{tab:NetworkModelRobust}}
\end{table}

\begin{figure}[h!]
\centering
\includegraphics[width=0.925\textwidth]{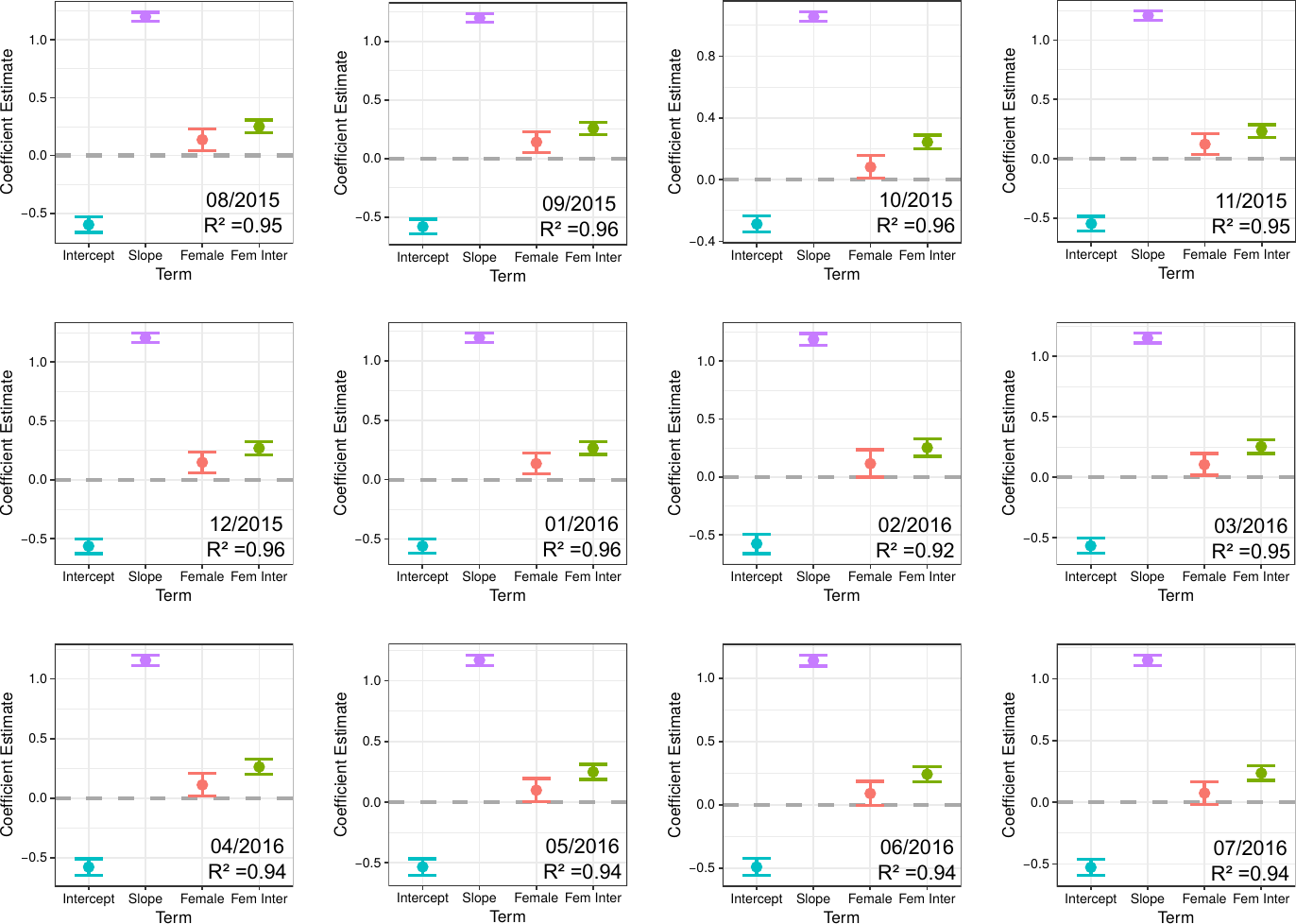}
\caption{Results of repetition of the fit of the network externalities model for 12 months between 2015 and 2016. The results are qualitatively stable across months. \label{fig:networkmodelMonth}} 
\end{figure}

As with previous models, we evaluated the model of network externalities over twelve months following our initial measurement. Fig. \ref{fig:networkmodelMonth} reports the overall results, showing no relevant decrease in $R^2$ and generally the same result, where the parameter  $\alpha_F$ is significantly larger than zero and the parameter $\alpha$ is significantly larger than one.

\begin{figure}[h!]
\centering
\includegraphics[width=0.925\textwidth]{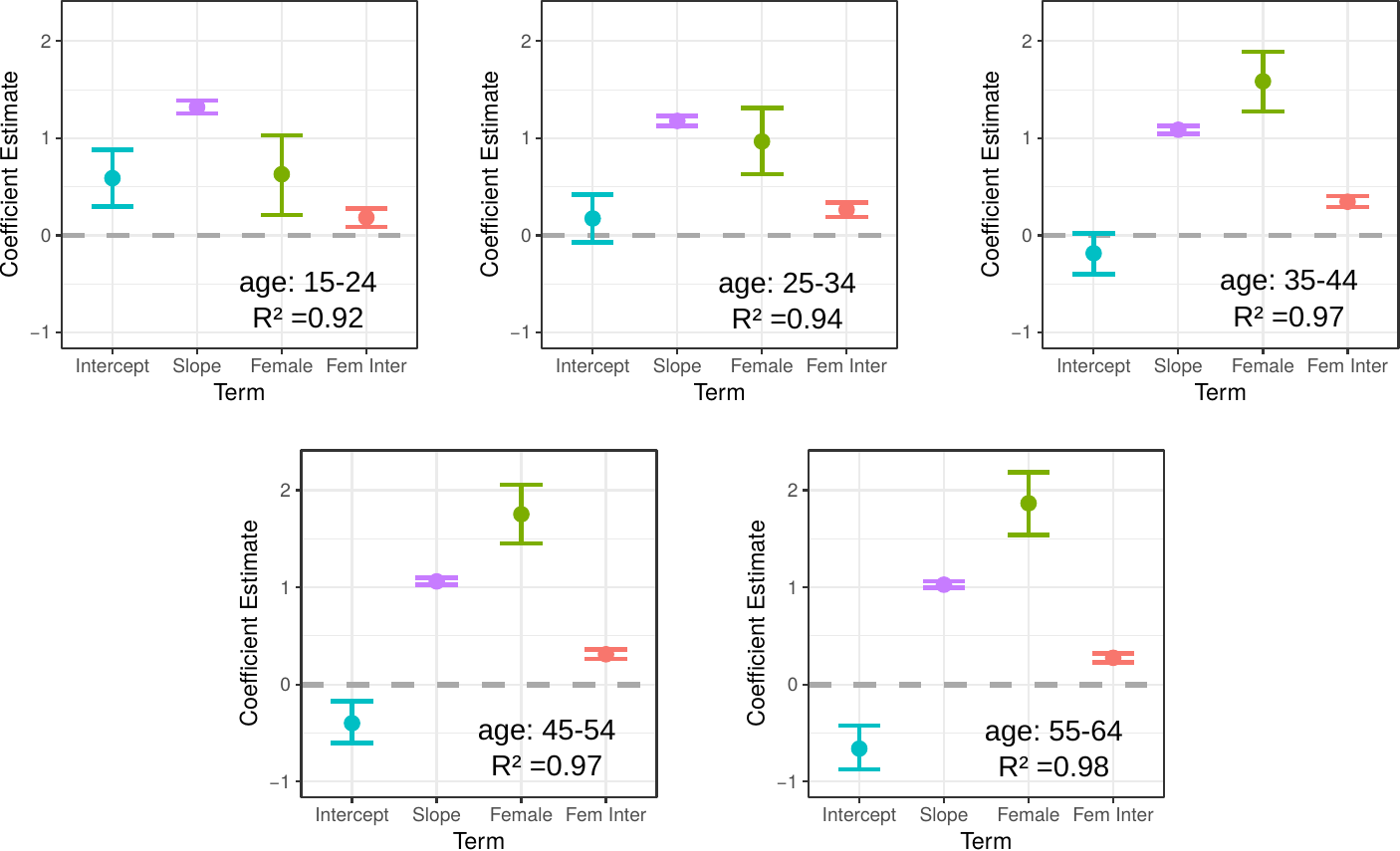}
\caption{Results of repetition of the fit of the network effects model for age segments. \label{fig:networkmodelAges}} 
\end{figure}

We stratified the analysis, fitting the network externalities model using calculations of the FGD using only data from a set of age categories. Fig. \ref{fig:networkmodelAges} shows the model results, evidencing that the female intercept, which measures the surplus of the exponent for female users, is positive and significant for all age categories.

\clearpage

\section*{Supplementary Text 4 - Gender equality changes}

Table \ref{tab:VIF2} reports the Variance Inflation Factors of the variables in the model of economic gender equality changes and in the model of changes of FGD. All factors are low enough to discard multi-collinearity.

\begin{table}[h]
\centering
\begin{tabular}{|c c | c c |}
\hline
\multicolumn{2}{|c|}{$\Delta Eco$ Model} & \multicolumn{2}{c|}{$\Delta FGD$ Model} \\
\hline
Variable & VIF & Variable & VIF\\
\hline
FGD Rank & 1.833& FGD & 1.66 \\
Economic Gender Equality 2015 & 1.276 & Rank Economic Gender Equality 2015 & 1.15 \\
GDP Rank & 1.505  & GDP Rank & 1.493 \\
\hline
\end{tabular}
\caption{Variance Inflation Factors of independent variables in the economic gender equality changes model. \label{tab:VIF2}}
\end{table}

Table \ref{tab:ModelComparison} presents the detailed results of both models of changes.
Before fitting, we rescaled the ranked variables to have a value between zero and one to allow a better comparison of their relationships, controlling for autocorrelation by including the unranked value of the variable in the previous year.
The results of Table \ref{tab:ModelComparison} are confirmed by ANOVA tests of the FGD rank in the $\Delta Eco$ model $F=10.195, p<0.01$,  and the non-significant result for the Eco rank in the $\Delta FGD$ model $F=0.003, p>0.9$.

\begin{table}[h]
\centering
\begin{tabular}{|c c c c | c c c c|}
\hline
\multicolumn{4}{|c|}{$\Delta Eco$ Model} & \multicolumn{4}{c|}{$\Delta FGD$ Model} \\
\hline
Term & Estimate & s.e & p-value & Term & Estimate & S.e & p-value\\
\hline
Intercept & -0.011 & 0.015 & 0.437 & Intercept & 0.019 & 0.010 & 0.069\\
FGD Rank & \bf 0.039 & 0.01 & $<0.01$ & FGD & -0.002 & 0.011 & 0.848 \\
Eco 	 & \bf -0.061 & 0.021 & $<0.005$ & Rank Eco & -0.001 & 0.015 & 0.94 \\
GDP Rank & \bf 0.052 & 0.01 & $<10^{-6}$ & GDP Rank & 0.007 & 0.015 & 0.65 \\
\hline
\multicolumn{2}{|c}{Multiple $R^2$} & \multicolumn{2}{c|}{0.1501} & \multicolumn{2}{c}{Multiple $R^2$} & \multicolumn{2}{c|}{0.0009} \\
\hline
\end{tabular}
\caption{Results of robust regression for the model of changes in economic gender equality and of changes in the FGD.\label{tab:ModelComparison}}
\end{table}

\begin{figure}[h]
\centering
\includegraphics[width=0.95\textwidth]{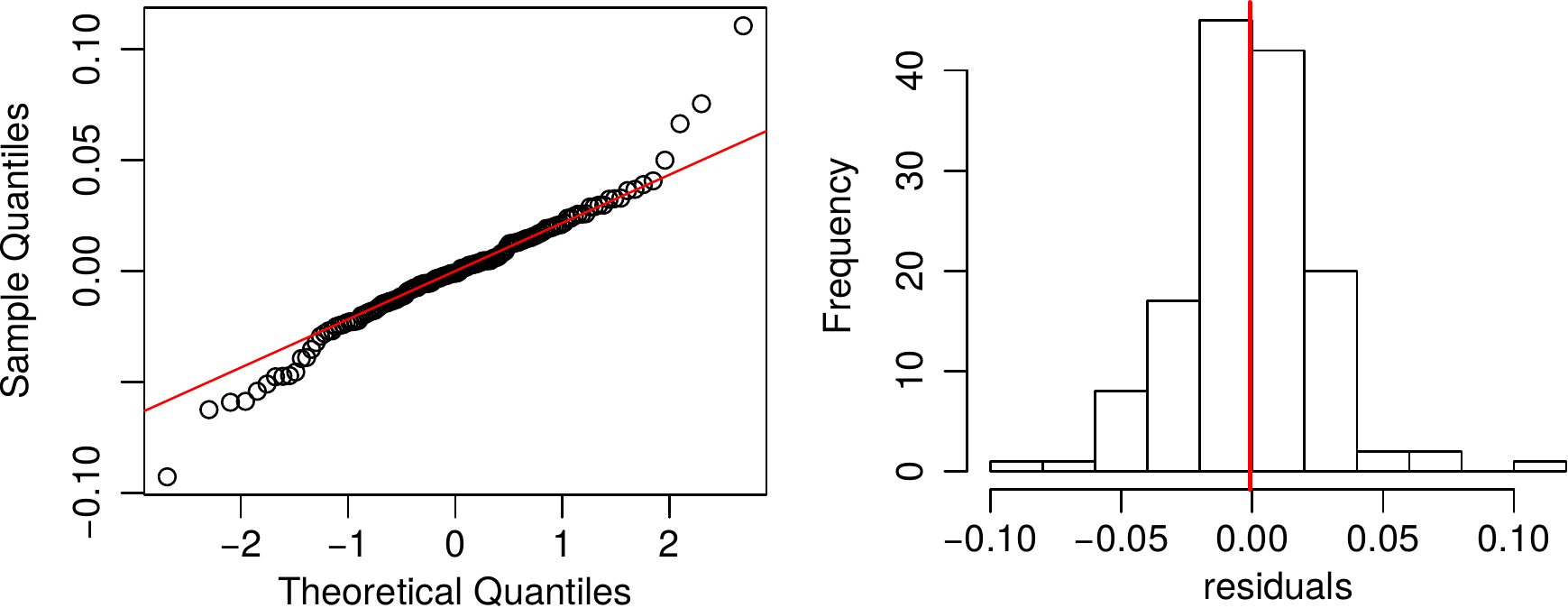}
\caption{Analysis of residuals of the economic gender equality changes model. The left panel shows the normal Q-Q plot
of residuals of the model, and the right panel their histogram. Some minor deviations from normality can be observed in both.
\label{fig:ChangeResiduals}} 
\end{figure}

The residuals of the model of changes in economic gender equality are distributed close to normality, as shown on Fig. \ref{fig:ChangeResiduals}, with a significant Shapiro-Wilk statistic of $0.97$ and only some small deviations from normality at the tails. Residuals are uncorrelated with all independent variables and do not show signs of heteroskedasticity.

We tested the robustness of the positive association between FGD and $\Delta Eco$ in two new fits including the same controls as for the FGD model (VIF of all factors below 5). The results are shown on Table \ref{tab:ModelsControl}, evidencing that the observed association between FGD and $\Delta Eco$ is robust to other socio-economic indicators, and to other possible confounds such as Facebook penetration or mean user age.

\begin{table}[h]
\centering
\begin{tabular}{|c | c c c | c c c|}
\hline
Term & Estimate & s.e & p-value & Estimate & s.e & p-value\\
\hline
Intercept & 0.007690 & 0.018282 & 0.67473 & 0.006294 & 0.017968 & 0.72667 \\
FGD Rank & \bf 0.024313 & 0.010963 & $<0.05$ & \bf 0.026402 & 0.010911 &  $<0.05$ \\
Eco & \bf -0.072524 & 0.024797 & $<0.01$ & \bf -0.075975 &  0.024367&  $<0.01$ \\
Ineq Rank & \bf -0.016436 & 0.007985 & $<0.05$ & -0.013190 & 0.007922 & 0.09833\\
Pop Rank & \bf 0.017917 & 0.008364 & $<0.05$ & \bf 0.019372 & 0.008090 & $<0.05$ \\
Mean Age Rank & 0.005872 & 0.012132 & 0.62919 & 0.006279 & 0.011781  & 0.59494\\
FB Penetration Rank & 0.015300 & 0.013533 & 0.26029&  0.015453& 0.012387& 0.21442 \\
GDP Rank & 0.024484 & 0.016641 & 0.14359 &  & & \\
Internet Penetration Rank &  &  & &0.024763 & 0.015389& 0.11000\\
\hline
Multiple $R^2$ & \multicolumn{3}{|c|}{0.2009}&\multicolumn{3}{c|}{0.1966} \\
\hline
\end{tabular}
\caption{Results of the $\Delta Eco$ model including additional controls.\label{tab:ModelsControl}}
\end{table}

We further tested the possible role of other equality indices in the relationship between FGD and $\Delta Eco$.
We added all other three gender equality scores as controls (VIF below 5), and repeated the fit.
The result, shown on Table \ref{tab:ModelsControlEqu} shows that the positive association between FGD and $\Delta Eco$ is robust to the possible effect of other kinds of gender inequality.

\begin{table}[h]
\centering
\begin{tabular}{|c | c c c |}
\hline
Term & Estimate & s.e & p-value \\
\hline
Intercept & -0.029304 & 0.021736 & 0.179917 \\
FGD Rank  & \bf  0.039560 & 0.016828 & $<0.05$ \\
Economic Score       &  -0.043693 & 0.025742 & 0.091983 \\
GDP rank & \bf 0.045928  & 0.012187 & $<0.001$\\
Education Score rank & 0.001291  & 0.013318 & 0.922924 \\
Political Score rank & 0.014899 & 0.009621 & 0.123862 \\
Health Score rank & 0.004209  & 0.009031 & 0.641953 \\
\hline
N & 139 & Multiple $R^2$ & 0.171 \\
\hline
\end{tabular}
\caption{Results of the $\Delta Eco$ model including controls for other gender equality indices.\label{tab:ModelsControlEqu}}
\end{table}

\newpage

We tested whether the association between FGD and $\Delta Eco$ could be explained by general cultural differences.
We combined our dataset with Hofstede's cultural dimensions: Power Distance Index (PDI), Individualism (IDV), Uncertainty Avoidance Index (UAI), and Masculinity (MAS) (VIF below 5).  This limits the analysis to a set of 66 countries common to both datasets, with results reported on Table \ref{tab:ModelsControlHof}. The positive association between FGD and $\Delta Eco$ is still significant, suggesting that the relationship between both variables goes beyond what Hofstede's model captures in terms of culture.

\begin{table}[h]
\centering
\begin{tabular}{|c | c c c |}
\hline
Term & Estimate & s.e & p-value \\
\hline
Intercept & 0.04743 & 0.03102 & 0.1316\\
FGD Rank & \bf 0.03792 & 0.01764 & $<0.05$ \\
Eco & \bf -0.1243& 0.03779 & $<0.01$ \\
PDI & 3.217$* 10^{-4}$ &1.759$* 10^{-4}$  &0.0725 \\
IDV & 1.754$* 10^{-4}$  & -1.708$* 10^{-4}$ & 0.3088\\
MAS &  -2.490$* 10^{-4}$ & 1.546$* 10^{-4}$ & 0.1126\\
UAI & 3.148$* 10^{-5}$ &  1.366$* 10^{-4}$& 0.8185 \\
\hline
N & 66 & Multiple $R^2$ & 0.362 \\
\hline
\end{tabular}
\caption{Results of the $\Delta Eco$ model including controls for cultural dimensions.\label{tab:ModelsControlHof}}
\end{table}

\begin{figure}[h!]
\centering
\includegraphics[width=0.95\textwidth]{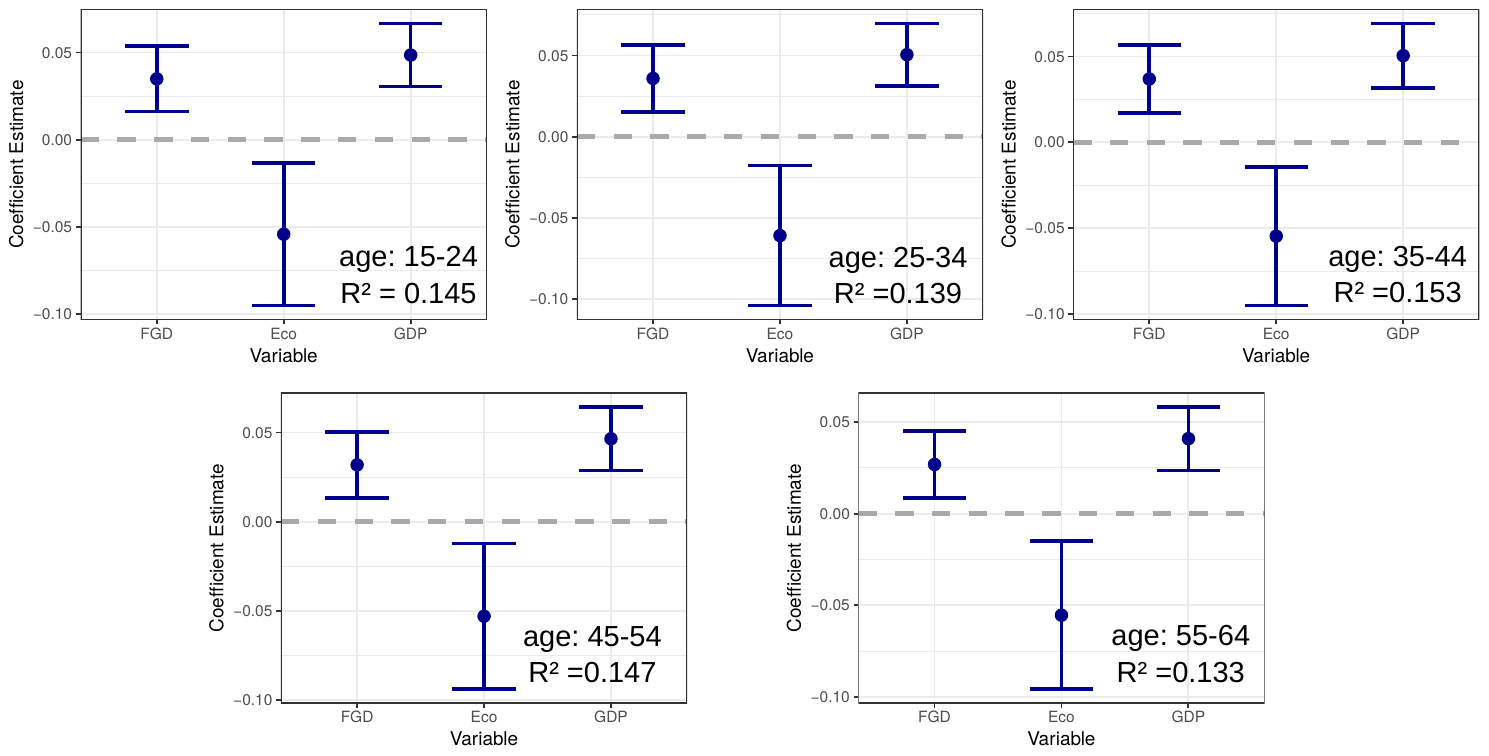}
\caption{Replication of the model stratifying for different age groups. \label{fig:Model3Ages}} 
\end{figure}

Following the same methodology as for previous models, we stratified our analysis with FGD measured only in a variety of age groups. The results are shown on Fig. \ref{fig:Model3Ages}, revealing the positive and significant role of FGD in the model for all age segments. 

\newpage

Our data offers the opportunity to measure the role of FGD in the changes of other gender equality measures, but the indices for Political and Health gender equality have negligible changes between 2015 and 2016. For that reason we can only evaluate the role of FGD for changes in Education gender equality. We find no significant effect of FGD, as reported on Table \ref{tab:EduModel}.

\begin{table}[h]
\centering
\begin{tabular}{|c | c c c |}
\hline
Term & Estimate & s.e & p-value \\
\hline
Intercept & \bf 0.0613210 & 0.0613210 & $<10^{-7}$\\
FGD rank & -0.0004546 &  0.0023720  & 0.848 \\
Edu & -0.0606702 & 0.0099782 & $<10^{-6}$  \\
GDP rank & \bf -0.0012934 & 0.0022238 &  0.562     $<0.01$ \\
\hline
N & 139 & Multiple $R^2$ & 0.08 \\
\hline
\end{tabular}
\caption{Results of the $\Delta Edu$ model.\label{tab:EduModel}}
\end{table}

\end{document}